\documentclass[11pt]{article}
\usepackage{amsmath}
\usepackage{latexsym}

\allowdisplaybreaks

\renewcommand{\baselinestretch}{1.2}
\newcommand{\be}{\begin{equation}}
\newcommand{\ee}{\end{equation}}
\newcommand{\bea}{\begin{eqnarray}}
\newcommand{\eea}{\end{eqnarray}}
\newcommand{\nn}{\nonumber} 
\newcommand{\cL}{\mathcal{L}}
\newcommand{\cD}{\mathcal{D}}

\newcommand{\cV}{\mathcal{V}}
\newcommand{\cF}{\mathcal{F}}
\newcommand{\cFm}{F}

\newcommand{\Mm}{{\bf M}}

\newcommand{\I}{\mathrm{i}}

\newcommand{\mh}{\hat{\mu}}

\def\slash#1{\rlap{\hbox{$\mskip 1 mu /$}}#1}      
\def\Slash#1{\rlap{\hbox{$\mskip 3 mu /$}}#1}      

\begin{document}
\begin{titlepage}
\begin{flushright} \small
 ITP-UU-06/26 \\ SPIN-06/22 \\ hep-th/0606148
\end{flushright}
\bigskip

\begin{center}
 {\LARGE\bfseries Off-shell N=2 tensor supermultiplets}
\\[10mm]
\textbf{B. de Wit and F. Saueressig}\\[5mm]
{\em Institute for Theoretical Physics} and {\em Spinoza
  Institute,\\ Utrecht University, Utrecht, The Netherlands}\\[3mm] 
{\tt  B.deWit@phys.uu.nl}\;,\; 
{\tt F.S.Saueressig@phys.uu.nl}
\end{center}

\vspace{5mm}

\bigskip

\centerline{\bfseries Abstract} 
\medskip
\noindent
A multiplet calculus is presented for an arbitrary number $n$ of $N=2$
tensor supermultiplets. For rigid supersymmetry the known couplings
are reproduced. In the superconformal case the target spaces
parametrized by the scalar fields are cones over $(3n-1)$-dimensional
spaces encoded in homogeneous $\mathrm{SU}(2)$ invariant potentials,
subject to certain constraints.  The coupling to conformal
supergravity enables the derivation of a large class of supergravity
Lagrangians with vector and tensor multiplets and hypermultiplets.
Dualizing the tensor fields into scalars leads to hypermultiplets with
hyperk\"ahler or quaternion-K\"ahler target spaces with at least $n$
abelian isometries. It is demonstrated how to use the calculus for the
construction of Lagrangians containing higher-derivative couplings of
tensor multiplets. For the application of the c-map between vector and
tensor supermultiplets to Lagrangians with higher-order derivatives,
an off-shell version of this map is proposed. Various other
implications of the results are discussed. As an example an elegant
derivation of the classification of 4-dimensional quaternion-K\"ahler
manifolds with two commuting isometries is given. \medskip
\end{titlepage}
\section{Introduction}
\label{sec:intro}
The importance of off-shell methods for the construction of
supersymmetric Lagrangians is well known. For $N=2$ supersymmetry in
four space-time dimensions the most relevant off-shell supermultiplets
are the Weyl, the vector and the tensor supermultiplet. The Weyl
supermultiplet comprises the fields of conformal supergravity, whereas
the other two multiplets play the role of matter multiplets. The
hypermultiplet does not constitute an off-shell multiplet, unless one
introduces an infinite number of fields.  This paper deals with $N=2$
tensor supermultiplets whose off-shell formulation has a long history.
In \cite{dWvH} the multiplet emerged as a submultiplet of off-shell
$N=2$ supergravity. Its transformation rules in a general
superconformal background were presented in \cite{dWvHVP} and a
locally superconformally invariant Lagrangian for a single tensor
multiplet was written down in \cite{deWit:1982na}. The latter enabled
the derivation of an alternative minimal off-shell formulation of
$N=2$ supergravity.

In four space-time dimensions it is possible to dualize a rank-two
tensor gauge field into a scalar field. In this way actions of tensor
supermultiplets lead to corresponding supersymmetric actions for
hypermultiplets. The resulting hypermultiplet target space will then
have a group of abelian isometries induced by the gauge invariance of
the tensor fields. In the case of rigid supersymmetry the
hypermultiplets parametrize a hyperk\"ahler space. In
\cite{Lindstrom:1983rt,Hitchin:1986ea} the $N=1$ superspace
formulation was used to classify, upon dualization, $4n$-dimensional
hyperk\"ahler metrics with $n$ abelian isometries. The Lagrangians are
encoded in terms of a function subject to certain partial differential
equations, which can be elegantly written in terms of a contour
integral depending on the tensor multiplet scalars.  Furthermore a
first $N=2$ superspace formulation was presented in
\cite{Karlhede:1984} in which this contour integral played a central
role.

In the context of local $N=2$ supersymmetry one is interested in
superconformal tensor multiplets. The scalar fields then parametrize
target spaces which are cones over a $(3n-1)$-dimensional space. When
coupling these supermultiplets, together with at least one vector
multiplet and possible hypermultiplets, to conformal supergravity, the
resulting theory is gauge equivalent to Poincar\'e supergravity
coupled to matter fields.  In this gauge equivalence the number of
matter multiplets is reduced by two. This is so because part of the
components belonging to the two, so-called compensating,
supermultiplets correspond to superconformal gauge degrees of freedom.
Upon gauge fixing the remaining components of these multiplets combine
with the fields of the Weyl multiplet to constitute an off-shell
multiplet of Poincar\'e supergravity. There is a certain freedom in
choosing compensator multiplets, which leads to different off-shell
versions of Poincar\'e supergravity.  The more conventional one
employs a compensating vector multiplet and a hypermultiplet, but the
hypermultiplet can be replaced by a compensating tensor multiplet.
These two choices do in certain cases lead to the same theory as the
tensor fields can be dualized to scalar fields in which case the
hypermultiplet target space becomes a quaternion-K\"ahler space.
However, the dualization affects the off-shell supersymmetry
structure.

When dualizing superconformal Lagrangians of tensor multiplets one
obtains $4n$-dimensional hyperk\"ahler cones \cite{dWKV:1999}. The
latter are cones over $(4n-1)$-dimensional 3-Sasakian spaces, which in
turn are $\mathrm{Sp}(1)$ fibrations of $(4n-4)$-dimensional
quaternion-K\"ahler spaces. In this context the gauge-fixing of the
compensating degrees of freedom is known as a superconformal quotient
and this quotient was extensively studied in \cite{deWit:2001dj}. The
hyperk\"ahler cones are encoded in so-called hyperk\"ahler potentials
and it turns out that there exits a similar real function for
superconformal tensor multiplets that is homogeneous and
$\mathrm{SU}(2)$ invariant. Just like the function exploited in
\cite{Lindstrom:1983rt,Hitchin:1986ea} it is subject to a set of
partial differential equations. When applied to a single tensor
supermultiplet acting as a compensator (in addition to a compensator
vector supermultiplet), one recovers the results of
\cite{deWit:1982na} for pure supergravity with a tensor gauge field
and local $\mathrm{U}(1)$ invariance. In this setting the tensor field
does not describe dynamic degrees of freedom. For two tensor
multiplets one finds pure supergravity with an additional matter
multiplet, which contains two scalar and two tensor fields. Upon
dualization of the tensor fields one obtains supergravity coupled to a
single hypermultiplet whose target space defines a 4-dimensional
quaternion-K\"ahler space.  Solving the differential equations for the
$\mathrm{SU}(2)$ invariant potential of the tensor formulation, one
elegantly reproduces the general classification of the corresponding
4-dimensional metrics with two commuting Killing fields
\cite{CP:2001}. They include the metric of the so-called universal
hypermultiplet as a special case.

We should stress here that the above discussion is based on off-shell
supermultiplets. When one is just interested in supersymmetric
Lagrangians involving tensor fields, there are many more
possibilities, as one can always dualize tensor gauge fields into
scalar fields and, provided there are abelian isometries, vice versa.
For a general discussion of $N=2$ supersymmetric Lagrangians involving
tensor and scalar fields, we refer to \cite{Theis:2003jj}. Naturally,
these general Lagrangians are not encoded in a single function, unlike
the Lagrangians derived through the superconformal quotient, but there
are good reasons to believe that they can be derived from the same
formalism by a series of dualizations \cite{deWit:2001dj}.

The superconformal quotient for tensor supermultiplets was extensively
discussed in \cite{deWit:2001dj} without paying attention to the
details of their supergravity couplings. The first topic of this paper
is therefore to extend the results of \cite{deWit:1982na} to an
arbitrary number of tensor supermultiplets.  In the case of rigid
supersymmetry, the results of this paper are completely in accord with
\cite{Hitchin:1986ea}. It turns out that the coupling to conformal
supergravity is straightforward in the present framework. The results
can be used in the context of string compactifications where tensor
fields arise naturally. Some of the results of this paper have already
been exploited to derive string-loop corrected hypermultiplet metrics
for type-II string theory compactified on a generic Calabi-Yau
threefold \cite{RSV}. Our work also has some overlap with, for
example, that of \cite{ZG} where dimensional reductions of
five-dimensional supergravity theories are studied.  For general
gaugings the situation is less clear.  It is known that magnetic
background fluxes generically require the presence of tensor fields,
which, however, acquire non-trivial mass terms
\cite{Louis:2002ny,italians,dWST}.  Whether or not these tensor fields
are in some way related to the tensor fields that are discussed here,
is yet an open issue.

The results of this paper also enable the construction of
higher-derivative actions for tensor supermultiplets. These actions
contain terms of fourth order in space-time derivatives. We will
demonstrate this by presenting one non-trivial example of such an
action for a single tensor supermultiplet, encoded in a single
function subject to differential constraints. To couple such an action
to supergravity is straightforward and one has an additional option of
including independent couplings with the Weyl multiplet or with vector
multiplets in the form of a chiral background \cite{deWit:1996ix}. We
intend to give a more complete presentation of these higher-derivative
couplings elsewhere.

Vector supermultiplets can also have higher-derivative couplings. Also
here we distinguish between vector multiplet couplings with the Weyl
multiplet through a chiral background, and actions which contain {\it
  ab initio} higher-derivatives of the vector multiplet components
themselves.  The former are the ones relevant for the topological
string \cite{Bershadsky:1993ta} and have played an important role in
the comparison between microscopic and macroscopic black hole entropy
\cite{LopesCardoso:1998wt}.  The latter are of the type studied, for
example, in \cite{dWGR}. All these higher-order actions will
undoubtedly contribute to the Wald entropy \cite{Wald:1993nt}, which
was crucial for obtaining agreement between microscopic and macroscopic
black hole entropy at the subleading level in the limit of large
charges.

It is clearly of interest to investigate on a par the
higher-derivative couplings for both tensor and vector
supermultiplets, as those are expected to be related by the so-called
c-map \cite{CFG}.  Conventionally, the c-map is applied on the basis
of actions that are at most quadratic in space-time derivatives
\cite{FS,dWVP-V,RVV,Cortes:2005uq}. In this way classical tensor (and thus
hypermultiplet) moduli spaces that appear in compactifications of
type-II strings can be determined from vector moduli spaces, as a
result of T-duality. When considering actions with
higher-order derivatives, also derivatives of auxiliary fields appear.
Therefore we also study the definition of c-map for full off-shell
supermultiplets, independent of the actions considered. The application of
the c-map to higher-order derivative couplings was discussed in
\cite{Antoniadis:1994,Berkovits:1995cb} and in a recent paper \cite{RVV}. 

This paper is organized as follows. In section~\ref{sec:rigid-tensors}
we discuss the tensor supermultiplets in the context of rigid
supersymmetry. Following and extending the results of
\cite{deWit:1982na}, we construct composite chiral multiplets in terms
of tensor multiplet components. Subsequently we proceed to derive
invariant actions. Furthermore we show how superconformally invariant
actions are encoded in terms of a homogeneous $\mathrm{SU}(2)$
invariant potential, similar to the hyperk\"ahler potentials for
superconformal hypermultiplet Lagrangians. In
section~\ref{sec:c-map-higer-derivative} we analyze the off-shell
version of the c-map between vector and tensor multiplets and we
present a nontrivial example of a supersymmetric action for a tensor
supermultiplet involving higher-order derivatives. In
section~\ref{sec:cf-sg-coupling} we consider the coupling of tensor
multiplets to conformal supergravity.  In section \ref{sec:Poincare}
we discuss the superconformal quotient for Lagrangians involving
tensor and vector multiplets and hypermultiplets to obtain Poincar\'e
supergravity theories with tensor multiplets. To demonstrate the
virtues of our formulation we consider the case of two tensor
multiplets and evaluate the differential equations for the
$\mathrm{SU}(2)$ invariant potential of the tensor formulation to
obtain the classification of the corresponding 4-dimensional selfdual
Einstein metrics with two commuting Killing fields.
Finally some details of the superconformal calculus are
presented in an appendix.

\section{Rigid tensor multiplet couplings}
\label{sec:rigid-tensors}
The $N=2$ tensor multiplet can be realized off-shell in a general
superconformal background. In this section we consider the case of
rigid supersymmetry in flat Minkowski space.  The tensor
supermultiplet is described in terms of a tensor gauge field
$E_{\mu\nu}$, an ${\rm SU(2)}$ triplet of scalar fields $L^{ij}$, a
doublet of Majorana spinors $\varphi^i$ and an auxiliary complex
scalar $G$. The supersymmetry transformation rules can be written as
follows \cite{dWvHVP},
\begin{eqnarray}
  \label{eq:3.1}
\delta L_{ij}&=& 2\,\bar\epsilon_{(i}\varphi_{j)}
+2 \,\varepsilon_{ik}\varepsilon_{jl}\, \bar\epsilon^{(k}\varphi^{l)}
\,,\nn\\  
\delta\varphi^i&=& \,\slash{\partial} L^{ij} \,\epsilon_j +
\varepsilon^{ij}\,\Slash{E}\,\epsilon_j - G\,\epsilon^i \,,\nn\\ 
\delta G &=& -2 \bar\epsilon_i \,\slash{\partial}\varphi^i \,,\nn\\
\delta E_{\mu\nu} &=& i\bar\epsilon^i\gamma_{\mu\nu}
\varphi^j\,\varepsilon_{ij} - i\bar\epsilon_i\gamma_{\mu\nu}
\varphi_j\,\varepsilon^{ij} \,,
\end{eqnarray}
where (anti)symmetrization is always defined with unit strength
(unlike in \cite{deWit:1982na}). Gamma matrices
$\gamma^{\mu\nu\cdots}$ with multiple indices denote antisymmetrized
products of gamma matrices in the usual fashion. We recall that
$\epsilon^i$ and $\varphi^i$ are positive-chirality spinors whose
negative-chirality counterparts are denoted by $\epsilon_i$ and
$\varphi_i$, respectively. Furthermore,  $E^\mu =
\frac{1}{2}\mathrm{i} \varepsilon^{\mu\nu\rho\sigma} \partial_\nu
E_{\rho\sigma}$ is the invariant field strength of the tensor field. 
The scalar field $L_{ij}$ satisfies a reality
constraint, $L^{ij}= \varepsilon^{ik}\varepsilon^{jl}\,L_{kl}$.
Complex conjugation is effected by raising and lowering of ${\rm
  SU(2)}$ indices, $i,j,k,\ldots$. Throughout this paper we use
Pauli-K\"all\'en metric conventions.

\subsection{Composite reduced chiral supermultiplets}
\label{sec:reduced-chiral-multiplet}
Supersymmetric Lagrangians with at most two space-time derivatives can
be constructed by making use of the observation that a tensor
multiplet can couple linearly to a reduced chiral multiplet. The
latter supermultiplet comprises a complex scalar $X$, an antisymmetric
tensor $F_{\mu\nu}$, a (negative-chirality) spinor doublet $\Omega^i$
and a triplet of auxiliary scalars $Y^{ij}$. Its supersymmetry
transformations are as follows,
\begin{eqnarray}
  \label{eq:3.2}
  \delta X &=& \bar\epsilon^i \Omega_i\,,\nn\\
\delta \Omega_i &=& 2\,\slash{\partial} X\,\epsilon_i + \tfrac1{2}
\varepsilon_{ij}\, 
\gamma^{\mu\nu} F_{\mu\nu}\,\epsilon^j + Y_{ij}\,\epsilon^j \,, \nn \\
  \delta F^-_{\mu\nu} &=& \tfrac1{2} \bar \epsilon_i
  \slash{\partial}\gamma_{\mu\nu} \Omega_j\, \varepsilon^{ij} 
  - \tfrac1{2} \bar \epsilon^i \gamma_{\mu\nu}\slash{\partial} \Omega^j
    \,\varepsilon_{ij} \,,\nn\\ 
\delta Y_{ij} &=& 2\,\bar\epsilon_{(i}\, \slash{\partial} \Omega_{j)}
+ 2\,\varepsilon_{ik}\varepsilon_{jl}\,  \bar\epsilon^{(k}
  \,\slash{\partial} \Omega^{l)}  \,.
\end{eqnarray}
Here $F^-_{\mu\nu}$ is the antiselfdual component of the tensor
$F_{\mu\nu}$, whose complex conjugate equals $F^+_{\mu\nu}$. Because
we are dealing with a reduced chiral multiplet, $Y^{ij}$ 
satisfies a reality constraint, $Y^{ij}=
\varepsilon^{ik}\varepsilon^{jl}\,Y_{kl}$ and $F_{\mu\nu}$ satisfies a
Bianchi identity, $\partial_{[\mu}F_{\nu\rho]}=0$. The latter can be
solved (at least locally) so that $F_{\mu\nu}$ acquires the form  
$F_{\mu\nu}= \partial_\mu W_\nu -  \partial_\nu W_\mu$. The resulting
vector supermultiplet can then be completed by specifying the
transformation rule for $W_\mu$,
\begin{equation}
  \label{eq:vector-susy}
  \delta W_\mu= \bar \epsilon_i \gamma_\mu \Omega_j\, \varepsilon^{ij}
  + \bar \epsilon^i \gamma_\mu \Omega^j \,\varepsilon_{ij}\,. 
\end{equation}
As is well-known, there exists a non-abelian version of this multiplet
which will, however, not be needed in what follows.

The supersymmetric coupling of a tensor to a reduced chiral multiplet
takes the form,
\begin{eqnarray}
  \label{eq:vector-tensor}
  {\cal L} &=& X\,G + \bar X\,\bar G - \tfrac1{2} Y^{ij}\,L_{ij} + \bar
  \varphi^i\,\Omega_i + \bar \varphi_i\,\Omega^i - \tfrac1{4}i \,
  \varepsilon^{\mu\nu\rho\sigma} \,E_{\mu\nu} \,F_{\rho\sigma} \,.
\end{eqnarray}
This expression can be used to derive supersymmetric Lagrangians
for tensor multiplets, as was already demonstrated in
\cite{dWR,deWit:1982na}.  This derivation is based on the observation
that one can construct a reduced chiral multiplet from tensor
multiplet components. When substituting the
components of this composite multiplet into (\ref{eq:vector-tensor})
one obtains a supersymmetric Lagrangian for the tensor multiplet. 

In order to construct $n$ reduced chiral multiplets from $n$ tensor
multiplets, one must introduce a (real) function ${\cal F}_{I,J}(L)$
of the tensor multiplet scalars $L^{ijI}$, where we label the
$n$ tensor supermultiplets by upper indices $I, J,\ldots=
1,2,\ldots,n$. The reduced chiral multiplet to which each tensor
multiplet couples is then assigned a lower index $I$. The construction
starts from the lowest component of the chiral multiplet, which is
given by
\begin{equation}
  \label{eq:X}
  X_I = {\cal F}_{I,J}(L) \, \bar G^J +
 {\cal F}_{I,J,K}{}^{ij}(L)\,\bar\varphi_i{}^J\varphi_j{}^K \,, 
  \end{equation}
where 
\begin{equation}
  \label{eq:def-1-der}
  {\cal F}_{I,J,Kij}(L) = \frac{\partial {\cal F}_{I,J}(L)} {\partial
  L^{ijK}}\,. 
\end{equation}
We note that ${\cal F}_{I,J,Kij}$ satisfies the same reality
constraint as the fields $L_{ij}{}^I$. Hence its ${\rm SU}(2)$ indices
can be raised by complex conjugation, or alternatively, by contraction
with epsilon tensors. Such quantities define real ${\rm SU}(2)$
vectors and their products satisfy certain product relations which
reflect their decomposition in terms of irreducible ${\rm SU}(2)$
representations. We present two of them, which are used throughout this
paper. The products of two such real vectors, $L_{ij}$ and $K_{ij}$,
satisfy 
\begin{eqnarray}
  \label{eq:KL-product}
    K_{ik}\,L^{jk}+  K^{jk} \,L_{ik} &=& \delta^j_i \,  K_{kl}
    \,L^{kl}\,, \nn \\
    K_{ij}\,L_{kl} - K_{kl} \,L_{ij} &=& \varepsilon_{ik}\,
    \varepsilon^{mn} \,(K_{lm}
    \,L_{nj} +K_{jm}\,L_{nl}) \Big|_{(i,j)\, (k,l)}\,, 
\end{eqnarray}
where the right-hand side of the second equation is symmetrized in
$(i,j)$ and $(k,l)$. These identities can be used with
$K_{ij}$ or $L_{ij}$ equal to $L_{ij}{}^I$ or ${\cal F}_{I,J,Kij}$. 

To ensure that we are dealing with a chiral multiplet the
supersymmetry transformation of the composite field $X_I$ has to be of
the form \eqref{eq:3.2}. Up to terms cubic in the spinors
$\varphi_{i}{}^I$ this imposes that the derivative ${\cal
  F}_{I,J,Kij}$ must be {\it symmetric} in $(JK)$. The higher-order
spinor terms require a second condition, namely,
\begin{equation}
  \label{eq:cubic-spinor}
  {\cal F}_{I,J,K}{}^{ij}{}_{,L}{}^{kl}(L)
  \;(\bar\varphi_i{}^K\varphi_j{}^J)\; \varphi_k{}^L=0\,, 
  \end{equation}
where we defined 
\begin{equation}
  \label{eq:def-2-der}
  {\cal F}_{I,J,Kij,Lkl}(L) = \frac{\partial^2 {\cal F}_{I,J}(L)}
  {\partial L^{ijK}\,\partial L^{klL} }\;.
\end{equation}
When ${\cal F}_{I,J,Kij,Lkl}(L)$ is symmetric in $(jk)$ the cubic spinor
term (\ref{eq:cubic-spinor}) vanishes. It is therefore guaranteed that we
are dealing with a chiral multiplet once the following constraints are
satisfied by the function ${\cal F}_{I,J}$,
\begin{equation}
  \label{eq:chiral-constraints}
  {\cal F}_{I,J,Kij}= {\cal F}_{I,K,Jij}\,,\qquad 
   \varepsilon^{jk}\, {\cal F}_{I,J,Kij,Lkl}(L)=0\,.
\end{equation}
As it turns out these constraints also suffice to ensure that
we are dealing with a reduced chiral multiplet.  

The function ${\cal F}_{I,J}$ has no particular symmetry in $I$ and
$J$. From the constraints (\ref{eq:chiral-constraints}) it follows
that its derivatives with respect to the $L^{ijK}$ are independently
symmetric under the capital indices $J,K,\ldots$ and under the ${\rm
  SU}(2)$ indices $i,j,k,l,\ldots$. This motivates us to use an 
obvious notation ${\cal F}_{I,J_1\cdots J_{p+1} j_1\cdots j_{2p}}$ for
the $p$-th multiple derivative, which is symmetric in both the $p+1$
indices $\{J\}$ and in the $2\,p$ indices $\{j\}$.

Henceforth we assume that the conditions (\ref{eq:chiral-constraints})
are satisfied. From the variation of (\ref{eq:X}) we determine the
composite spinor field $\Omega_{iI}$ of the chiral multiplet,
\begin{eqnarray}
  \label{eq:3.4}
  \Omega_{i\,I} &=& -2 \,{\cal F}_{I,J}\,\slash{\partial}
  \varphi_i{}^J 
  + 2\, {\cal F}_{I,JKij} \, \bar G^J\, \varphi^{jK}  -  2\,
    {\cal F}_{I,JK}{}^{kl} \,(\,\slash{\partial} L_{ik}{}^J -\varepsilon_{ik}
    \Slash{E}^J ) \varphi_l{}^K    \nn  \\ 
 &&{}  
  + 2\, {\cal F}_{I,JKLij}{}^{kl}  \;\varphi^{jL}
    \,(\bar\varphi_k{}^J\varphi_l{}^K)  \,. 
\end{eqnarray} 
The supersymmetry variation of $\Omega_{i\,I}$ yields the expressions
for $Y_{ij\,I}$ and $F_{\mu\nu\,I}$, while all remaining variations
correctly recombine into the derivative $\partial_\mu X_I$. The
explicit expressions for the new fields read,
\begin{eqnarray}
  \label{eq:Y-F}
   Y_{ij \,I} &=& -2\,{\cal F}_{I,J}\, \partial^2 L_{ij}{}^J
   -2\,{\cal F}_{I,JKij} \,( \bar G^J\,G^K + E_{\mu}{}^J\,E^{\mu K}) \,,
  \nn\\ 
&&{}
  -2\, {\cal F}_{I,JK}{}^{kl}\,( \partial_\mu L_{ik}{}^J\,\partial^\mu
  L_{jl}{}^K + 2\,\varepsilon_{k(i}\, \partial_\mu L_{j)l}{}^J\,
  E^{\mu K} ) \nn\\ 
&&{} 
   -2 \,{\cal F}_{I,JKLij}{}^{kl} \;\bar\varphi_{k}{}^K\varphi_{l}{}^J \,
   G^L 
   -2 \,{\cal F}_{I,JKLijkl} \;\bar\varphi^{kK}\varphi^{lJ} \, \bar
   G^L    \nn\\ 
&&{} 
   + 4\,( {\cal F}_{I,JK m(i}\,\bar\varphi^{mJ} \,\slash{\partial}
     \varphi_{j)}{}^K + {\cal F}_{I,JK}{}^{m(k} \bar{\varphi}_m{}^J
   \slash{\partial} \varphi^{l)K} \, \varepsilon_{ik} \, \varepsilon_{jl})
   \nn\\ 
&&{} 
   + 4 \,{\cal F}_{I,JKLn(i}{}^{kl} \,\partial_\mu L_{j)k}{}^J
     \left(\bar{\varphi}^{nL} \gamma^\mu \varphi_{l}{}^K \right)  \nn\\ 
&&{}
   -4 \,{\cal F}_{I,JKLn(i}{}^{kl}\, \varepsilon_{j)k} \;\left(
   \bar{\varphi}^{nL} \, \Slash{E}^J \, \varphi_{l}{}^K \right) \nn\\
&&{}
   -2 \, {\cal F}_{I,JKLMijmn}{}^{kl} \, \bar\varphi_{k}{}^J
   \varphi_{l}{}^K \;\bar\varphi^{mL} \varphi^{nM}  \;,   
\nn\\ 
   F_{\mu\nu\,I} &=&  {}- 2\,{\cal F}_{I,JK}{}^{mn} \,\partial_{[\mu}
   L_{mk}{}^J \, \partial_{\nu]} L_{nl}{}^K \,\varepsilon^{kl} \nn\\  
   &&{}
   -4\,\partial_{[\mu} \Big({\cal F}_{I,J} \,
   E_{\nu]}{}^J + {\cal F}_{I,JKki}\, \bar \varphi
   ^{kJ}\gamma_{\nu]} \varphi_{j}{}^K \,\varepsilon^{ij} \Big) \, .
\end{eqnarray} 
The results can be compared to the corresponding ones given in
\cite{deWit:1982na}.  In order that we are dealing with a single
reduced chiral superfield for given index $I$, it is important that
${\cal F}_{I,J}$ is a real function. This enables the use of
identities such as \eqref{eq:KL-product}. These identities and
(\ref{eq:chiral-constraints}) are used throughout the calculation. The
Bianchi identity holds for $F_{\mu\nu\,I}$, although the second term
proportional to $\partial_{[\mu} L\,\partial_{\nu]} L$ is somewhat
subtle. By virtue of (\ref{eq:chiral-constraints}) the contribution of
this term, $\partial_{[\mu} F_{\nu\rho]} \propto {\cal
  F}_{I,JKLijkl}\varepsilon_{mn}\, \partial_{[\mu} L^{ijJ}
\,\partial_\nu L^{kmK} \,\partial_{\rho]}L^{lnL}$, vanishes so that
$F_{\mu\nu}$ is closed. However, $F_{\mu\nu}$ is not exact in the
sense that it cannot be written as the curl of a manifestly ${\rm
  SU}(2)$ invariant quantity. We will exhibit this below.

Let us now discuss the constraints (\ref{eq:chiral-constraints}). To
analyze their implications, we decompose the field $L^{ijI}$ into a
real field $x^I$ and a complex field $v^I$ according to, 
\begin{equation}
  \label{eq:L-xv}
  L^{12\,I} \equiv \tfrac1{2} \mathrm{i}\,x^I\,, \qquad
  L^{11\,I}\equiv v^I\,,
\end{equation}
so that $L_{ij}^I L^{ij\,J}= \tfrac1{2} x^Ix^J+ 2\, v^{(I}\bar v^{J)}$. 
The constraints (\ref{eq:chiral-constraints}) then take the following
form,\footnote{
   Derivatives with respect to $L_{ij}{}^I$ are defined by
   $\frac{\partial}{\partial L_{ij}{}^J} \, L_{kl}{}^I = \frac{1}{2}
   \left( \, \delta_k^i \, \delta_l^j + \delta_l^i \, \delta_k^j
   \right) \, \delta^I_J$, so that $\delta L_{ij}{}^I
   \partial/\partial L_{ij}^I = \delta x^I \partial/\partial x^I
   +\delta v^I\partial/\partial v^I +\delta \bar v^I\partial/\partial \bar
   v^I$.} 
\begin{eqnarray}
  \label{eq:chiral-constr-2}
  &&
  \frac{\partial{\cal F}_{I,J}}{\partial x^K} =\frac{\partial
  {\cal F}_{I,K}}{\partial x^J}  \;, \qquad
  \frac{\partial{\cal F}_{I,J}}{\partial v^K} =\frac{\partial
  {\cal F}_{I,K}}{\partial v^J}  \;, \nn\\
  &&
  \frac{\partial^2{\cal F}_{I,J}}{\partial x^K\partial x^L} +
  \frac{\partial^2{\cal F}_{I,J}}{\partial v^K \partial \bar v^L} =0
  \;. 
\end{eqnarray}
The last equation, which simply follows from $\mathcal{F}_{I,JKL
ij}{}^{ij} =0$, contains the $\mathrm{SU}(2)$ invariant Laplacian,
\begin{equation}
  \label{eq:laplacian}
  \tfrac1{2}\,\varepsilon_{ik} \varepsilon_{jl}\, \frac{\partial^2}{\partial
  L_{ij}{}^I\,\partial L_{kl}{}^J} = \frac{\partial^2}{\partial
  x^I\,\partial x^J} + \frac{\partial^2}{\partial v^{(I} \,\partial
  \bar v^{J)}} \,.
\end{equation}
As a consequence of the first equation of (\ref{eq:chiral-constr-2}),
${\cal F}_{I,J}$ can be expressed as a derivative of a new function
${\cal F}_I$ which is, however, still constrained,
\begin{eqnarray}
  \label{eq:chiral-constr-3}
  &&
  {\cal F}_{I,J} = \frac{\partial {\cal F}_{I}}{\partial x^J} \;,
  \qquad
  \frac{\partial^2 {\cal F}_{I}}{\partial x^J\partial v^K} =
  \frac{\partial^2 {\cal F}_{I}}{\partial x^K\partial v^J}\;, \nn\\ 
  &&
  \frac{\partial^2{\cal F}_{I}}{\partial x^J\partial x^K} +
  \frac{\partial^2{\cal F}_{I}}{\partial v^J\partial\bar v^K} =0 \;.  
\end{eqnarray}
The last equation of (\ref{eq:chiral-constr-3}) was determined by
integrating the last equation of (\ref{eq:chiral-constr-2}) which
leaves a real function on the right-hand side that does not depend on
$x$. However, differentiation with respect to $v^L$ (or $\bar v^L$)
yields a function symmetric in $(J,L)$ (or $(K,L)$) which implies that
the right-hand side can be written as the $\partial^2/\partial
v^J\partial \bar v^K$ derivative of some function of $v$ and $\bar v$.
As ${\cal F}_I$ is defined up to an $x$-independent function, the
latter can be absorbed into ${\cal F}_I$.

With these results we can now exhibit that the expression for
$F_{\mu\nu\,I}$ given in (\ref{eq:Y-F}) takes indeed the form of a
curl, 
 \begin{equation}
  \label{eq:mono-potential}
  {\cal F}_{I,JK}{}^{ij} \, \partial_{[\mu} L_{ik}{}^J
  \,\partial_{\nu]} L_{jl}{}^K \,\varepsilon^{kl} =
  \mathrm{i}\,\partial_{[\mu}\Big( \frac{\partial
  \mathcal{F}_I}{\partial \bar v^J}\,\partial_{\nu]}\bar v^J- 
  \frac{\partial\mathcal{F}_I}{\partial v^J}\,\partial_{\nu]}v^J
  \Big)\,,   
\end{equation}
so that the Bianchi identity is manifestly satisfied. 

Let us close with two examples which lead to Lagrangians (constructed
according to the procedure outlined in the next subsection) that are
both dual to non-interacting hypermultiplets. One concerns the simple
example where $\mathcal{F}_{I,J}=\delta_{IJ}$ is $L$-independent. This
example trivially satisfies the constraints
(\ref{eq:chiral-constraints}). One possible expression for
$\mathcal{F}_I$ takes the form,
\begin{equation}
  \label{eq:F-cal-free}
  \mathcal{F}_I = x^I + c_{IJ}\,v^J + \bar c_{IJ}\,\bar v^J\,,
\end{equation}
with $c_{IJ}$ some complex constants. A second example is based on
the conformal tensor multiplet introduced in
\cite{deWit:1982na,Lindstrom:1983rt}, where $\cF_{I,J} =
\delta_{IJ}\,(L^I)^{-1}$ with $L^I =\sqrt{L_{ij}{}^I\, L^{ijI}}$, so
that, for $I,J,K,L$ equal,
\begin{equation}
   \label{eq:improved-tensor}
   {\cal F}_{I,JKij}(L) = - \frac{L_{ij}{}^{I}}{(L^I)^3}\,,\qquad  
  {\cal F}_{I,JKL ijkl}(L) = \frac{3\, L_{ij}{}^{I}\, L_{kl}{}^{I}
  +(L^I)^2\,   \varepsilon_{i(k}  \varepsilon_{l)j}}{(L^I)^5}\,,
\end{equation}
which satisfies the constraints (\ref{eq:chiral-constraints}). 
A corresponding expression for $\mathcal{F}_I$ is given by
\begin{equation}
  \label{eq:F-cal-impr}
  \mathcal{F}_I = \sqrt{2} \ln\Big[x^I  + \sqrt{x^Ix^I + 4 \,v^I \bar
  v^I}\Big] - \tfrac1{2}\sqrt{2} \ln\Big[4 \,v^I \bar v^I\Big]\,. 
\end{equation}

\subsection{Supersymmetric tensor multiplet actions}
\label{sec:tensor-actions}
We now proceed to give the rigidly supersymmetric tensor multiplet
Lagrangian obtained by substituting the composite fields \eqref{eq:X},
\eqref{eq:3.4} and \eqref{eq:Y-F} into the density formula
(\ref{eq:vector-tensor}). Up to total derivatives the Lagrangian
equals 
\begin{eqnarray}
  \label{eq:rigid-lagrangian}
  {\cal L}&=&{} 
  F_{IJ}\Big[-\tfrac12 
   \partial_\mu L_{ij}{}^I\,\partial^\mu L^{ijJ} +E_{\mu}^I\, E^{\mu J}
   - (\bar \varphi^{iI}  \,\slash{\partial}\varphi_{i}{}^J +  \bar
     \varphi_{i}{}^I \,\slash{\partial}\varphi^{iJ} )+ G^I\,\bar G^J
     \Big]   \nn\\ 
&&{}
  +\tfrac12 ie^{-1} \varepsilon^{\mu\nu\rho\sigma} \,F_{IJK}{}^{ij}
  \,E_{\mu\nu}^I\,  \, \partial_\rho L_{ik}{}^J\, \partial_\sigma
  L_{jl}{}^K\,\varepsilon^{kl} \nonumber\\  
&&{}
   - F_{IJK}{}^{ij} \Big[\bar{\varphi}^{kI} \,\slash{\partial}
     L_{jk}{}^J \, \varphi_{i}{}^K  - G^I
     \,\bar\varphi_{i}{}^J\varphi_{j}{}^K \Big]  
   \nn\\
&&{}
   - F_{IJK\,ij} \Big[ \bar{\varphi}_{k}{}^I \,\slash{\partial}  
     L^{jkJ} \, \varphi^{iK} - \bar G^I
     \,\bar\varphi^{iJ}\varphi^{jK} \Big]     \nn\\
&&{}
  + 2\, F_{IJK}{}^{ij} 
    \, \varepsilon_{ki} \, \bar{\varphi}^{kI} \, \Slash{E}^J \,
    \varphi_{j}{}^K    \nn\\
&&{} 
  + F_{IJKL\,ij}{}^{kl} 
  \,\bar\varphi_{k}{}^I \,\varphi_{l}{}^J 
    \;\bar\varphi^{iK}\varphi^{jL}  \,,
\end{eqnarray}
where 
\begin{eqnarray}
  \label{eq:der-F}
  F_{IJ} &=&  2\,{\cal F}_{(I,J)} + L^{ijK}\, {\cal F}_{K,IJ}{}_{ij}\,,
  \nonumber \\[2mm] 
  F_{IJK}{}^{ij} &=& 
  3\,{\cal F}_{(I,JK)}{}^{ij}+ L^{klL} {\cal F}_{L,IJKkl}{}^{ij} 
  = \frac{\partial F_{IJ}}{\partial L_{ij}{}^K} \,,  \nn\\
  F_{IJKL\,ij}{}^{kl} &=& 
  4\, {\cal F}_{(I,JKL)ij}{}^{kl} + L^{mnM} {\cal
  F}_{M,IJKLmnij}{}^{kl} 
    = \frac{\partial^2 F_{IJ}}{\partial L^{ijK}\partial L_{kl}{}^L} \,. 
\end{eqnarray}
We note that the tensor gauge field always appears in form of the
covariant field strength $E^\mu$, with the exception of the second
line proportional to $\varepsilon^{\mu\nu\rho\sigma}$. This term is
nevertheless invariant under tensor gauge transformations, up to a
total derivative, owing to the Bianchi identity satisfied by the
$L$-dependent terms. In the basis \eqref{eq:L-xv}, this term can be
rewritten in terms of the tensor field strength after partial
integration, as we shall discuss shortly ({\it c.f.}
(\ref{eq:mono-potential-2})).

The Lagrangian is encoded in the function $F_{IJ}$ and its
derivatives. Making use of  (\ref{eq:chiral-constr-3}), the functions
$F_{IJ}$ can be written as follows,
\begin{eqnarray}
  \label{eq:F}
  F_{IJ} &=&
  \frac{\partial{\cal F}_I}{\partial x^J}  + 
  \frac{\partial{\cal F}_J}{\partial x^I} + x^K
  \frac{\partial^2{\cal F}_{K}}{\partial x^I\partial x^J} + v^K
  \frac{\partial^2{\cal F}_{K}}{\partial x^I\partial v^J} + \bar v^K 
  \frac{\partial^2{\cal F}_{K}}{\partial x^I\partial \bar v^J} \;,
  \nn\\ 
  &=&
  \frac{\partial}{\partial x^I}\Big[{\cal F}_J  
   + x^K   \frac{\partial{\cal F}_{K}}{\partial x^J} + v^K
  \frac{\partial{\cal F}_{K}}{\partial v^J} + \bar v^K 
  \frac{\partial{\cal F}_{K}}{\partial \bar v^J}\Big]  \;. 
\end{eqnarray}
This expression is symmetric in $(I,J)$. Thus the terms inside the
bracket are equal to the $x^J$-derivative of another function.
Therefore $F_{IJ}$ can be written as the second $x$-derivative of some
unknown function $F(x,v,\bar v)$. Integrating (\ref{eq:F}) yields the
first derivative of $F$,
\begin{equation}
  \label{eq:eq:F-1}
   \frac{\partial
  F}{\partial x^J} =   {\cal F}_J  + 
   x^K   \frac{\partial{\cal F}_{K}}{\partial x^J} + v^K
  \frac{\partial{\cal F}_{K}}{\partial v^J} + \bar v^K 
  \frac{\partial{\cal F}_{K}}{\partial \bar v^J}\,,
\end{equation}
up to an $x$-independent function which we set to
zero. Subsequently we evaluate $\partial^2 F/\partial
v^I\partial x^J$ and establish its symmetry in
$(I,J)$ from (\ref{eq:chiral-constr-3}). Furthermore we verify that
\begin{equation}
  \label{eq:laplace+x}
  \frac{\partial}{\partial x^I}\,\Big[
  \frac{\partial^2 F}{\partial x^J\partial x^K}+ 
  \frac{\partial^2 F}{\partial v^J\partial\bar v^K}   \Big]  =0\,,
\end{equation}
making use again of (\ref{eq:chiral-constr-3}). By following the same
argument as below (\ref{eq:chiral-constr-3}), one then establishes the
existence of a function $F$ subject to the equations,
\begin{equation}
  \label{eq:F-constraints}
  \frac{\partial^2 F}{\partial x^I\partial v^J} =  \frac{\partial^2
  F}{\partial x^J\partial v^I}\;, \qquad  
  \frac{\partial^2 F}{\partial x^I\partial x^J} +
  \frac{\partial^2 F}{\partial v^I\partial\bar v^J} =0 \;.  
\end{equation}
The Lagrangian is thus encoded in functions $F(x,v,\bar v)$, with
\begin{equation}
  \label{eq:F=ddF}
  F_{IJ} = \frac{\partial^2 F}{\partial x^I \partial x^J} \,,
\end{equation}
and $F(x,v,\bar v)$ subject to the conditions
(\ref{eq:F-constraints}).  This result is entirely consistent with the
results derived in \cite{Lindstrom:1983rt,Hitchin:1986ea}, where it
was shown how to express the function $F(x,v,\bar v)$ in terms of a
contour integral.

Using the above relations we derive, along the same lines as in
(\ref{eq:mono-potential}), the relation,
 \begin{equation}
  \label{eq:mono-potential-2}
  F_{IJK}{}^{ij} \, \partial_{[\mu} L_{ik}{}^J
  \,\partial_{\nu]} L_{jl}{}^K \,\varepsilon^{kl} =
  \mathrm{i}\,\partial_{[\mu}\Big( \frac{\partial^2
  F}{\partial x^I\,\partial \bar v^J}\,\partial_{\nu]}\bar v^J- 
  \frac{\partial^2 F}{\partial x^I \,\partial v^J}\,\partial_{\nu]}v^J
  \Big)\,.
\end{equation}
This result is needed when dualizing the tensor fields to scalars. In
that case the supersymmetry is no longer realized off shell. One
introduces a new set of fields, $y_I$, which act as Lagrange
multipliers to impose the Bianchi identity on the tensor field
strength. Adding the term $y_I\, \partial_\mu  E^{\mu I}$
to the Lagrangian and integrating out the $E^{\mu I}$, one obtains an
action for hypermultiplets. A natural set of complex variables then
consists of the complex fields $v^I$ and $w_I$. The latter are defined
by \cite{deWit:2001dj}
\begin{equation}
  \label{eq:w-fields}
  w_I = \frac1{2}\Big(\mathrm{i} y_I + \frac{\partial F}{\partial x^I}
  \Big) \,.
\end{equation}
In terms of these fields the kinetic term of the scalar fields reads,
\begin{equation}
  \label{eq:hyperdual-sigma-model}
  \mathcal{L} = -F_{IJ} \,\partial_\mu v^I \partial^\mu \bar v^J -
  F^{IJ} \Big(\partial_\mu w_I - \frac{\partial^2F}{\partial x^I v^K}
  \,\partial_\mu v^K \Big) \Big
  (\partial^\mu \bar w_J - \frac{\partial^2F}{\partial   x^J \bar v^L}
  \,\partial^\mu \bar v^L 
  \Big)  \,,  
\end{equation}
where $F^{IJ}$ is the inverse of $F_{IJ}$. 

For completeness we present the functions $F(x,v,\bar v)$ corresponding to
the two examples (\ref{eq:F-cal-free}) and (\ref{eq:F-cal-impr}),
respectively,
\begin{eqnarray}
  \label{eq:ex-F}
  F(x,v, \bar v) &=& \sum_I \Big\{(x^I)^2 - 2\, v^I\bar v^I\Big\}
  \,, \nn\\
  F(x,v,\bar v) &=& \sqrt{2}\, \sum_I \Big\{x^I \ln \Big[x^I + 
  \sqrt{(x^I)^2 + 4\,v^I\bar v^I}\Big]  
  +  \tfrac1{2}(1-x^I) \ln\Big [4\, v^I\bar v^I\Big] \nonumber\\
  &&\hspace {12mm }-   \sqrt{(x^I)^2 +  4\,v^I\bar v^I}\,\Big\} \,. 
\end{eqnarray}

\subsection{Superconformal actions and tensor and hyperk\"ahler cones}
\label{sec:tensor-cones}
So far our analysis was completely general and we did not insist on
any additional invariance beyond $N=2$ supersymmetry. However, a tensor
supermultiplet can be assigned to a representation of the full $N=2$
superconformal algebra and the function ${\cal F}_{I,J}$ can be
chosen such that the composite chiral supermultiplet constitutes also
a superconformal representation. By substituting the superconformally
invariant composite chiral multiplets into the density formula these
symmetries carry over to the Lagrangian. The class of superconformal
actions is encoded by functions ${\cal F}_{I,J}$ that satisfy the
additional restriction,
\begin{equation}
  \label{eq:sc-tensor}
  {\cal F}_{I,JKik}\,L^{kjK} = -\tfrac1{2} \delta_i{}^j \,{\cal
  F}_{I,J}\,.  
\end{equation}
This condition, which will be derived in
section~\ref{sec:cf-sg-coupling}, implies that ${\cal F}_{I,J}$ is a
homogeneous function of the $L^{ijI}$ of degree $-1$ that is invariant
under the ${\rm SU}(2)$ R-symmetry.  It is easy to see that the
function $F_{IJ}$ that appears in the Lagrangian
(\ref{eq:rigid-lagrangian}), is thus also homogeneous of degree $-1$
and ${\rm SU}(2)$ invariant. Upon contraction with $L_{jm}{}^J$ one
proves another useful result,
\begin{equation}
  \label{eq:aux-terms}
  {\cal F}_{I,JK ij} \, L_{kl}{}^J \, L^{klK} = - {\cal F}_{I,J}
  \,L_{ij}{}^J \, , 
\end{equation}
which is needed later on. The same result applies to $F_{IJ}$. 

The constraint (\ref{eq:sc-tensor}) implies that the function ${\cal
  F}_I$ can be restricted to a homogeneous function of zeroth degree,
but, in general, it is only invariant under a $\mathrm{U}(1)$ subgroup
of the $\mathrm{SU}(2)$ R-symmetry.  The superconformal constraints on
the function $F(x,v,\bar v)$, which is a homogeneous function of
degree $+1$, were extensively analyzed in \cite{deWit:2001dj}. For
convenience, we summarize the conditions on the function $F_{IJ}$. In
the general case we have the constraints,
\begin{equation}
  \label{eq:F-constraint2}
  F_{IJK}{}^{ij} = F_{(IJK)}{}^{ij}\,, \qquad F_{IJKL}{}^{i[jk]l}
  =0\,. 
\end{equation}
For conformally invariant Lagrangians there is the additional
constraint,
\begin{equation}
  \label{eq:F-conf-constraints}
  F_{IJKik} \,L^{kj K} = - \tfrac1{2} \delta_i{}^j\, F_{IJ}\,. 
\end{equation}

When keeping the $x^I$ fixed, the subspace parametrized by the complex
fields $v^I$ is a K\"ahler space whose corresponding K\"ahler
potential is equal to the function $-F(x,v,\bar v)$. In the
conformally invariant case a similar potential exists for the target
space parametrized by the $L_{ij}{}^I$, which is defined by the
$\mathrm{SU}(2)$ invariant expression,
\begin{equation}
  \label{eq:tensor-potential}
  \chi_{\mathrm{tensor}}(L) = 2\, F_{IJ} \,L^{ij I}L_{ij}{}^J\,,
\end{equation}
and is a homogeneous function of degree $+1$. This potential is
closely related to the so-called hyperk\"ahler potential that plays a
similar role in the hypermultiplet case. To see this we first note
that its derivative with respect to $L^I$ is equal to the homothetic
vector,
\begin{equation}
  \label{eq:homothetic}
  \frac{\partial\chi_{\mathrm{tensor}}(L)}{\partial L_{ij}{}^I} = 2\, 
  F_{IJ}  L^{ij J} \,. 
\end{equation}
This vector generates the scale transformations on $L_{ij}{}^I$ with
scaling weight equal to 2. Furthermore we establish that the metric
$F_{IJ}$ is related to the second-order derivative of the potential,
according to
\begin{equation}
  \label{eq:chi-metric} 
    \varepsilon_{kl}\;
  \frac{\partial^2\chi_{\mathrm{tensor}}(L)}
  {\partial  L_{ik}{}^I\,\partial  L_{jl}{}^J}= 2\, F_{IJ}(L) 
  \;\varepsilon^{ij} \,.
\end{equation}
This implies that the $3n$-dimensional target space parametrized by
the $L_{ij}{}^I$ is a cone over a $(3n-1)$-dimensional space. The
potential $\chi_{\mathrm{tensor}}$ fully encodes the superconformal
theories of tensor supermultiplets. From it the function $F(x,v,\bar
v)$ can be determined by integration. In section \ref{sec:Poincare}
the role of $\chi_{\mathrm{tensor}}$ will be clarified further.

To elucidate the above, let us formulate it in terms of the variables
$v^I$, $\bar v^I$ and $x^I$. Using \eqref{eq:F=ddF} one establishes
the following identity,  
\begin{equation}
  \label{eq:tensor-potential-2}
  \chi_{\mathrm{tensor}}(L) =  F_{IJ}(x^I x^J + 4\, v^I \bar v^J) = -
  F(v,\bar v, x) + x^I \,\frac{\partial F(x,v,\bar v)}{\partial x^I} \,,
\end{equation}
where we made use of the various identities for derivatives of the
function $F(x,v,\bar v)$. The right-hand side of
\eqref{eq:tensor-potential-2} coincides with the expression for the
hyperk\"ahler potential given in \cite{deWit:2001dj} for the
hyperk\"ahler cones that one obtains upon dualizing the tensor fields
to scalars. Here the $x^I$ are expressed in terms of the coordinates
$w_I+\bar w_I$ given in \eqref{eq:w-fields}. Obviously the
hyperk\"ahler potential $\chi_{\mathrm{hyper}}(w_I,\bar w_I,v^I,\bar
v^I)$ and the function $F(x,v,\bar v)$ are related by a Legendre
transform.

The formalism of this paper makes it straightforward to incorporate
the coupling of tensor supermultiplets to conformal supergravity. In
\cite{deWit:1982na} this was demonstrated for a single tensor
supermultiplet and in section~\ref{sec:cf-sg-coupling} we will
generalize this result to $n$ tensor supermultiplets. Before turning
to this topic we first discuss a number of other features in the next
section.

\section{Off-shell c-map and higher-derivative actions}
\label{sec:c-map-higer-derivative}
\setcounter{equation}{0}
We have already stressed the importance of dealing with off-shell
supermultiplets which offer many technical advantages. In the first
subsection~\ref{sec:c-map} we will illustrate this once more by
introducing the c-map between off-shell tensor and vector
supermultiplets, outside the context of specific supersymmetric
actions. The fact that the c-map can be defined in this way is crucial
for its application to higher-derivative actions, where the existence
of an off-shell formulation is almost imperative. Without off-shell
multiplets higher-derivative actions can only be constructed by an
infinite series of iterations. Therefore we also briefly consider the
construction of higher-derivative couplings of tensor supermultiplets
in a second subsection~\ref{sec:high-deriv-actions}. The coupling to
supergravity will be the subject of later sections, but  we will
already present the extra bosonic terms that are generated in the
coupling to supergravity.

\subsection{The off-shell c-map}
\label{sec:c-map}
As is well known, four-dimensional vector- and hypermultiplet actions
are related to each other via the so-called c-map. Originally
\cite{CFG} this map was constructed by performing a dimensional
reduction of the four-dimensional action on a circle and dualizing the
three-dimensional vector field to a scalar. Because these operations
do not affect supersymmetry, the vector multiplets are converted into
hypermultiplets, so that one will be dealing with two hypermultiplet
sectors. Interchanging the two sectors and lifting back to a
four-dimensional action (assuming that the initial hypermultiplet
sector is itself in the image of the c-map) yields the desired map
between vector- and hypermultiplet sectors in four dimensions.

A more natural way to define the c-map is by comparing a dimensionally
reduced vector supermultiplet to a dimensionally reduced tensor
supermultiplet. Indeed it is immediately clear that there exists a close
relationship between the off-shell degrees of freedom. When reducing
on a circle in the 3-direction, the space-time coordinate vector
$x^\mu$ decomposes into a three-dimensional space-time vector
$x^{\hat\mu}$ ($\hat \mu= 0,1,2$) and a single coordinate $x^3$ which
will be shrunk to a point so that the fields become $x^3$-independent.
In this way the bosonic fields of the tensor multiplet decompose
according to, 
\begin{equation}
  \label{eq:tensor-red}
  \Big\{L_{ij}, E^{\mu}, G, \bar G  \Big\} \longrightarrow  \Big\{L_{ij},
  E^{\hat\mu}, E^3, G ,\bar G \Big\} \;,
\end{equation}
where $E^{\hat \mu}$ is a divergence-free vector field. Likewise the
bosonic fields of the (abelian) vector multiplet decompose according to,
\begin{equation}
  \label{eq:vector-red}
  \Big\{X, \bar X, F_{\mu\nu}, Y^{ij} \Big\} \longrightarrow  \Big\{X,
  \bar X, F_{\hat\mu 3},  F_{\hat\mu\hat\nu}, Y^{ij} \Big\} \sim 
\Big\{X,\bar X, W_3, F^{\hat\mu}, Y^{ij} \Big\}\;.
\end{equation}
In the last step we made use of the Bianchi identity satisfied by
$F_{\mu\nu}$, which implies that $F_{\hat\mu\hat\nu}$ is equivalent to
a divergence-free three-vector $F^{\hat\mu}= \tfrac1{4}\mathrm{i}
\varepsilon^{\hat\mu\hat\nu\hat\rho} F_{\hat\nu\hat\rho}$ and that
$F_{\hat\mu3}$ can be written as the derivative of a scalar field
$W_3$.  Hence the two multiplets are very similar. They both comprise
a single divergence-free vector, three physical scalars and three
auxiliary scalars, and they have the same number of fermionic degrees
of freedom. Both divergence-free vectors can be expressed in terms of a
vector potential which coincides (up to a gauge transformation) with
$E_{\hat\mu 3}$ and $W_{\hat\mu}$, respectively. 

The relation between the two supermultiplets becomes even more
striking upon realizing that the R-symmetry group, the relativistic
automorphism group of the supersymmetry algebra, which equals
$\mathrm{SU}(2)\times\mathrm{U}(1)$ in four space-time dimensions, is
extended to $\mathrm{SU}(2)\times\mathrm{SU}(2)$ in three space-time
dimensions.  Since the action of the $\mathrm{U}(1)$ subgroup is known
on the four-dimensional fields, it is not difficult to deduce the
representation content of the fields in three dimensions.  Obviously,
the fermionic fields must transform according to the $(2,2)$
representation of $\mathrm{SU}(2)\times\mathrm{SU}(2)$, while the
triplets $Y_{ij}$ and $L_{ij}$ transform according to the $(3,1)$
representation. Finally the triplets $\{X,\bar X,W_3\}$ and $\{G,\bar
G,E^3\}$ must transform according to the $(1,3)$ representation.
Obviously the two off-shell multiplets are the same and only differ in
their identification with the $\mathrm{SU}(2)$ factors of the
R-symmetry group.
 
The above conclusions are confirmed by an evaluation of the
supersymmetry transformation rules in the three-dimensional context,
following \cite{DeJaegher:1997ka}. First we define gamma matrices
$\hat \gamma^{\hat\mu}$ that are appropriate for the
three-dimensional theory, 
\begin{equation}
  \label{eq:gamma-3}
  \hat\gamma^{\hat\mu}= \gamma^{\hat \mu} \,\tilde\gamma\,,
\end{equation}
where $\tilde\gamma= - \mathrm{i} \gamma^3\gamma^5$ is an hermitean
matrix whose square is equal to the identity matrix. The product
$\hat\gamma^0\hat\gamma^1\hat\gamma^2$ is proportional to the identity
matrix. The hermitean matrices $\tilde\gamma$, $\gamma^3$ and
$\gamma^5$ commute with the $\hat\gamma^{\hat\mu}$ and consitute the
generators of an $\mathrm{su}(2)$ algebra that is related to the
second $\mathrm{SU}(2)$ factor of the R-symmetry group in three
dimensions. Obviously, $\mathrm{i}\gamma^5$, the $\mathrm{U}(1)$
R-symmetry generator of the four-dimensional theory is contained. The
second set of $\mathrm{SU}(2)$ transformations mixes spinors of
different chirality.  On the supersymmetry parameters with
(anti)chiral components $\epsilon^i$ ($\epsilon_i$), the `hidden'
$\mathrm{SU}(2)$ transformations act according to
\cite{DeJaegher:1997ka},
\begin{equation}
  \label{eq:second-su2}
  \delta\epsilon^i =- \tfrac1{2} \mathrm{i} \alpha\,\epsilon^i +
  \tfrac1{2} \beta \,\varepsilon^{ij} \gamma^3\,\epsilon_j\,,\qquad
  \delta\epsilon_i = \tfrac1{2} \mathrm{i} \alpha\,\epsilon_i +
  \tfrac1{2}  \bar \beta\,\varepsilon_{ij} \gamma^3\,\epsilon^j\,,
\end{equation}
where $\alpha$ is a real parameter associated with the
chiral $\mathrm{U}(1)$ R-symmetry in four dimensions and $\beta$ is
complex. It is straightforward to verify that the above
transformations generate a group $\mathrm{SU}(2)$ that commutes with
the four-dimensional $\mathrm{SU}(2)$ R-symmetry group. 

Now we present the three-dimensional supersymmetry transformations
upon the reduction to three space-time dimensions, which readily
follow from (\ref{eq:3.1}), (\ref{eq:3.2}) and (\ref{eq:vector-susy}),
and identify the R-symmetry transformations.  The result for the
tensor multiplet reads as follows\footnote{
  Note that the Dirac conjugate of a spinor involves the matrix
  $\gamma^0$. Therefore there is a relative factor $\tilde{\gamma}$
  between the three- and four-dimensional Dirac conjugates, and
  correspondingly between the two charge conjugation matrices.
  Therefore the three-dimensional charge conjugation matrix $\hat C$
  satisfies the following identities,
  \begin{equation}
    \hat C \hat\gamma^{\mh} \hat C^{-1} = - \hat\gamma^{\mh {\rm T}}
    \, , \quad  
    \hat C \gamma^{3} \hat C^{-1} =  \gamma^{3 {\rm T}}
    \,,\quad 
    \hat C \tilde{\gamma} \hat C^{-1} =  \tilde{\gamma}^{{\rm T}}
    \,,\quad 
    \hat C \gamma^{5} \hat C^{-1} =  - \gamma^{5 {\rm T}} \,,
    \quad \hat C^{\rm T} = - \hat C \, . \nonumber
\end{equation}
} 
\begin{equation}
  \label{eq:TM3d}
\begin{split}
  \delta L_{ij} = & \, 
  2\,\mathrm{i} \,\bar\epsilon_{(i} \gamma^{3} \varphi_{j)}
  -2\,\mathrm{i} \,\varepsilon_{ik} \,\varepsilon_{jl} \, 
  \bar\epsilon^{(k}  \gamma^{3}  \varphi^{l)} \,, \\
    \delta E_{\hat\mu 3} = & 
    \, \mathrm{i} \bar{\epsilon}^i \, \hat\gamma_{\mh} \, \gamma^{3}
    \, \varphi^j \, \varepsilon_{ij} 
    - \mathrm{i} \bar{\epsilon}_i \, \hat\gamma_{\hat\mu} \, \gamma^{3} \,
      \varphi_j \, \varepsilon^{ij} \, , \\ 
  \delta\varphi^i = & \,
  \mathrm{i} \, \hat{\slash{\partial}} L^{ij} \, \gamma^{3} \, \epsilon_j 
    + \mathrm{i} \varepsilon^{ij} \, E^{\hat\mu} \hat\gamma_{\hat\mu}
      \gamma^{3} \, 
      \epsilon_j  
    + \varepsilon^{ij} E^3 \, \gamma^{3} \epsilon_j - G\,\epsilon^i
      \,,\\   
    \delta E^3 = & 
    \, - \bar{\epsilon}^i  \gamma^{3} \hat{\slash{\partial}} \varphi^j \,
    \varepsilon_{ij} - \bar{\epsilon}_i \gamma^{3}
    \hat{\slash{\partial}}     \, \varphi_j \, \varepsilon^{ij} \, , \\ 
    \delta G =& \, -2 \bar\epsilon_i \,\hat{\slash{\partial}}\varphi^i
    \,,  
\end{split}
\end{equation}
where $\hat{\slash\partial}\equiv \hat\gamma^{\hat \mu}
\partial_{\hat\mu}$. The correct R-symmetry transformations can now be
identified by adopting $\mathrm{SU}(2)$ transformations for the
fermion fields $\varphi^i$, such that $\delta L_{ij}$ and $\delta
E_{\hat\mu 3}$ remain invariant under the combined transformations of
the fermions and the supersymmetry parameters. This leads to
\begin{equation}
  \label{eq:ssu2-varphi}
  \delta\varphi^i = \tfrac1{2} \mathrm{i} \alpha\,\varphi^i -
  \tfrac1{2} \bar \beta \,\varepsilon^{ij} \gamma^3\,\varphi_j\,,\qquad
  \delta\varphi_i = -\tfrac1{2} \mathrm{i} \alpha\,\varphi_i -
  \tfrac1{2} \beta\,\varepsilon_{ij} \gamma^3\,\varphi^j\,.
\end{equation}
The above transformations indeed generate the $\mathrm{SU}(2)$ group.
It is then straightforward to establish that under this particular
R-symmetry subgroup, the fields $G$, $\bar G$ and $E^3$ transform
according to the vector representation,
\begin{equation}
  \label{eq:ssu2-GGE}
  \delta G = \mathrm{i} \alpha \,G - \bar{\beta} \, E^3\,,
  \qquad
  \delta E^3 = \tfrac12\beta\,G + \tfrac12\bar\beta\,\bar G\,.
\end{equation}

Likewise, the dimensionally reduced supersymmetry transformations of
the vector supermultiplet read,
\begin{equation}
  \label{eq:VM3d}
\begin{split}
  \delta X = & \, - \mathrm{i} \, \bar{\epsilon}^i \, \gamma^3
  \, \Omega_i \, , \\ 
  \delta W_3 = & 
  \, \mathrm{i} \, \bar{\epsilon}_i \, \Omega_j \, \varepsilon^{ij} 
  - \, \mathrm{i} \, \bar{\epsilon}^i \, \Omega^{j} \,
    \varepsilon_{ij} \, , \\ 
  \delta W_{\hat\mu} = & \, 
  \bar{\epsilon}_i \hat\gamma_{\hat\mu} \, \Omega_{j} \,  
  \varepsilon^{ij}
  + \bar{\epsilon}^i \hat\gamma_{\hat\mu} \, \Omega^j \,
    \varepsilon_{ij} \, , \\ 
  \delta \Omega_i = & 
  \, 2\mathrm{i} \, \hat{\slash{\partial}} X \, \gamma^3 \, \epsilon_i 
  + \mathrm{i} \hat{\slash{\partial}} W_3 \, \varepsilon_{ij} \, \epsilon^j
  - \varepsilon_{ij} \, F^{\hat\mu}\hat\gamma_{\hat\mu} 
  \epsilon^j + Y_{ij} \, \epsilon^j \, , \\
  \delta Y_{ij} = & \, 
  2 \, \bar{\epsilon}_{(i}  \hat{\slash{\partial}} \, \Omega_{j)} 
  + 2 \, \varepsilon_{ik} \, \varepsilon_{jl}
   \, \bar{\epsilon}^{(k}  \hat{\slash{\partial}} \, \Omega^{l)} \, .  
\end{split}
\end{equation}
where $F^{\hat\mu}= \tfrac1{2} \mathrm{i}
\varepsilon^{\hat\mu\hat\nu\hat\rho} \partial_{\hat\nu} W_{\hat\rho}$.
The R-symmetry transformations of $\Omega^i$ follow from the
invariance of $\delta W_{\hat\mu}$ and $\delta Y_{ij}$ under the
combined transformations on the spinors. This time we find
\begin{equation}
  \label{eq:ssu2-Omega}
  \delta\Omega^i = \tfrac1{2} \mathrm{i} \alpha\,\Omega^i +
  \tfrac1{2} \bar\beta \,\varepsilon^{ij} \gamma^3\,\Omega_j\,,\qquad 
  \delta\Omega_i = -\tfrac1{2} \mathrm{i} \alpha\,\Omega_i +
  \tfrac1{2}  \beta\,\varepsilon_{ij} \gamma^3\,\Omega^j\,.
\end{equation}
which also correctly generates the $\mathrm{SU}(2)$ group associated
with (\ref{eq:second-su2}). It then follows that the fields $X$, $\bar
X$ and $W_3$ transform under the R-symmetry group according to the
vector representation, 
\begin{equation}
  \label{eq:ssu2-XXW}
  \delta X = - \mathrm{i} \alpha\, X +\tfrac1{2} \beta\, W_3\,,
  \quad
  \delta W_3 = -\beta\,\bar X - \bar \beta \, X \,.
\end{equation}

The above results enable the identification of the vector and tensor
multiplet components, up to an overall constant and an
$\mathrm{SU}(2)$ transformation that identifies the $\mathrm{U}(1)$
subgroup.  To see this we write the spinor quantities for the tensor
multiplet in a different basis. Following \cite{DeJaegher:1997ka} we
first write the supersymmetry parameters in a basis where the `hidden'
$\mathrm{SU}(2)$ factor of the R-symmetry becomes manifest,
\begin{eqnarray}
  \label{eq:spinor-rel}
  \epsilon^{+} = &  \tfrac12 \, \sqrt{2} \, \gamma^3 \left( \epsilon_1 -
   \mathrm{i} \epsilon_2 \right)  \; , \qquad
  \epsilon^{-} = &  \tfrac12 \, \sqrt{2} \,  \left( \epsilon^1
   -\mathrm{i} \epsilon^2 \right)  \, ,  \nn \\
  \epsilon_{+} = &  \tfrac12 \, \sqrt{2} \, \gamma^3 \left( \epsilon^1 +
  \mathrm{i}\epsilon^2 \right) \; , \qquad
  \epsilon_{-} = &  \tfrac12 \, \sqrt{2}  \, \left(\epsilon_1 +
  \mathrm{i} \epsilon_2 \right) \, .
\end{eqnarray}
To appreciate this choice of basis we note that the $\mathrm{SU}(2)$
transformations (\ref{eq:second-su2}) read 
\begin{equation}
  \label{eq:su2-epsilon-pm}
  \delta\epsilon^+= \tfrac12\mathrm{i} ( \alpha \,\epsilon^+ + \bar
  \beta \, \epsilon^-) \,. 
\end{equation}
Note that $\epsilon^{\pm}$ and $\epsilon_{\pm}$ are related through
charge conjugation. Likewise we write the tensor multiplet spinors as, 
\begin{eqnarray}
    \label{eq:c-map1}
    \varphi^{+} = & - \tfrac12 \, \sqrt{2} \, \gamma^3 \, \left( \varphi^1
    + \mathrm{i}\varphi^2 \right)\; , \qquad
   \varphi^{-} = & - \tfrac12 \, \sqrt{2} \, \left( \varphi_1 +
   \mathrm{i}\varphi_2 \right)  \,, \nn \\ 
   \varphi_{+} = & - \tfrac12 \, \sqrt{2} \, \gamma^3 \, \left( \varphi_1
   - \mathrm{i} \varphi_2 \right)\;, \qquad
   \varphi_{-} = & - \tfrac12 \, \sqrt{2} \, \left( \varphi^1
   -\mathrm{i} \varphi^2 \right) \,,                            
\end{eqnarray}
where the relevance of this basis follows from
\begin{equation}
  \label{eq:su2-varphi-pm}
  \delta\varphi^+= \tfrac12\mathrm{i} ( \alpha \,\varphi^+ + \bar
  \beta \, \varphi^-) \,.   
\end{equation}
In this basis the supersymmetry transformations of the tensor
multiplet can be compared directly to those of the vector multiplet
components, where we identify the spinor fields
$(\varphi^+,\varphi^-)$ with the spinor fields $(\Omega^1,\Omega^2)$
of the vector multiplet. This establishes the c-map for the bosonic
degrees of freedom,
\begin{equation}
  \label{eq:c-map2}
\begin{split}
  & L_{12} = \mathrm{i} (X+\bar X)\,, \quad  L_{11}= W_3 + X-\bar X \,,
  \quad  L_{22}= W_3 - X+ \bar X\,, \\ 
  & E_{\hat\mu 3} = W_{\hat\mu} \, , \\ 
  & G = Y_{22} \, , \quad \bar{G}= Y_{11}  \, , \quad  E^3 = \mathrm{i}
  Y_{12}  \, . 
\end{split}
\end{equation}

\subsection{On higher-derivative actions}
\label{sec:high-deriv-actions}
The expressions for the composite chiral supermultiplet can also be
used to construct actions with higher-derivative couplings. For
instance, we can start from the simple $N=2$ supersymmetric Lagrangian
for a single vector multiplet,
\begin{equation}
  \label{eq:vector>tensor}
 {\cal L} \propto  \vert \partial_\mu X\vert^2 + \tfrac1{8} \,
 F_{\mu\nu}{}^2 + \tfrac12 \bar{\Omega}^i \, \slash{\partial} \,
 \Omega_i -   \tfrac{1}{8} \, \vert Y_{ij}\vert^2  ,  
\end{equation}
and substitute the expressions for the composite components $X$,
$F_{\mu\nu}$, $\Omega_i$ and $Y_{ij}$ in terms of the tensor multiplet
components. These are encoded in a function $\mathcal{F}(L)$ subject
to the constraint,
\begin{equation}
  \label{eq:laplace-F}
  \frac{ \partial^2\mathcal{F}(L)}{\partial L^{ij}\,\partial L_{ij}} = 0
  \,. 
\end{equation}
This constraint enables one to show that the action depends only on a
single function $\mathcal{H}(L) = [\mathcal{F}(L)]^2$ which is no
longer subject to constraints. To demonstrate this we present the
bosonic terms,
\begin{equation}
  \label{eq:ho-lagrangian}
  \begin{split}
    \mathcal{L} = & \, \mathcal{H} \Big[  
      - \tfrac12 \vert\partial^2 L_{ij}\vert^2
  +2 \partial_{[\mu} E_{\nu]} \, \partial^{[\mu} E^{\nu]}+\vert
    \partial_\mu G \vert^2  \Big] \, \\
   & \, + \mathcal{H}^{ij} \Big[ 
   \left( \partial_{[\mu} L_{ik} \partial_{\nu]} L_{jl} \varepsilon^{kl}
   \right) \partial^\mu E^\nu  
   + 2 \left( \partial_{[\mu} L_{ij} E_{\nu]} \left( \partial^\mu E^\nu
  \right)  \right) 
  - \tfrac12 E^2 \partial^2 L_{ij} 
  - \vert G \vert^2 \, \partial^2 L_{ij} \\
  & \qquad \qquad 
  - \varepsilon_{ik} \big( E^\mu \partial_\mu L_{jl} \big) \big(
  \partial^2 L^{kl} \big) 
  - \tfrac12 \big( \partial_\mu L_{ik} \, \partial^\mu L_{jl} \big) \,
  \partial^2 L^{kl}  \Big] \\
& \, 
   - \tfrac12 \mathcal{H}^{ij,kl} \Big[ \big( \partial_\mu L_{ik} \,
    \partial^\mu L_{jl} \big) \vert G \vert^2  
   + \tfrac12 \varepsilon_{ik} \varepsilon_{jl} \big( \vert G \vert^2 +
   E^2 \big)^2 
   -2 \varepsilon_{ik} \big(E^\mu \partial_\mu L_{jl} \big) \big( \vert G
   \vert^2 + E^2 \big) \\ 
   & \, \qquad \qquad 
   -2 \varepsilon_{ik} \big( \partial_\mu L_{lp} \partial^\mu L^{np}
   \big) E^\nu \partial_\nu L_{jn} 
   - \varepsilon_{ik} \varepsilon_{lm} E^\mu E^\nu \big( \partial_\nu
   L^{mn} \big) \big( \partial_\mu L_{jn} \big) \\ 
   & \, \qquad \qquad
   -  \varepsilon_{ik} \varepsilon_{jm} E^2 \big( \partial_\mu L_{ln}
   \partial^\mu L^{mn}\big)  
   + \tfrac12 \, \varepsilon_{ik} \, \partial_\mu L_{jm} \, \partial^\mu
  L^{pq} \, \partial_\nu L_{nq} \, \partial^\nu L_{lp} \,
  \varepsilon^{mn}  
\Big] \, ,
\end{split}
\end{equation}
where 
\begin{equation}
  \label{eq:H-der}
  \mathcal{H}^{ij} = \frac{\partial\mathcal{H}}{\partial
  L_{ij}}\;,\qquad    \mathcal{H}^{ij,kl}  =
  \frac{\partial^2\mathcal{H}}{\partial L_{ij}\,\partial L_{kl}}\;. 
\end{equation}

Let us make a few comments at this point. First of all, consider the
linear combination of the free tensor multiplet Lagrangians
\eqref{eq:rigid-lagrangian} and \eqref{eq:ho-lagrangian},
\begin{eqnarray}
  \label{eq:free-higher-order}
  \mathcal{L} &=&{} 
   -\tfrac12 
   \vert \partial_\mu L_{ij}\vert^2  +E_{\mu}\, E^{\mu}
   - (\bar \varphi^{i}  \,\slash{\partial}\varphi_{i}{} +  \bar
     \varphi_{i}\,\slash{\partial}\varphi^{i} )+ \vert G\vert^2 
     \nn\\ 
     &&{}
     +M^{-2}\,\Big[ 
      - \tfrac12 \vert\partial^2 L_{ij} \vert^2 
  + 2 \partial_{[\mu} E_{\nu]} \, \partial^{[\mu} E^{\nu]}+ (\partial^2\bar
  \varphi^{i}  \,\slash{\partial}\varphi_{i}{} +  \partial^2\bar 
     \varphi_{i}\,\slash{\partial}\varphi^{i} ) + \vert
  \partial_\mu G \vert^2   \Big] \,,
\end{eqnarray}
where $M$ is a mass parameter.  This action describes a free massless
tensor multiplet and a {\it massive} vector supermultiplet, as can be
shown by analyzing the corresponding equations of motion. The massive
multiplet corresponds to negative metric states. All of this is in
accord with standard off-shell counting arguments.

Another comment concerns the R-symmetry. Lagrangians that are at most
quadratic in derivatives are always invariant under one of the factors
of the R-symmetry group, but not necessarily under both factors. For
instance, the two-derivative action for vector multiplets is always
invariant under the $\mathrm{SU}(2)$ R-symmetry subgroup but not
necessarily under the $\mathrm{U}(1)$ factor. For the tensor
multiplets the situation is precisely the reverse. In this respect the
Lagrangians that depend quartically on derivatives are different as
they can potentially break both factors of the R-symmetry group. 

Although it is in principle possible to convert the tensor field to a
scalar field by a duality transformation, the fact that the Lagrangian
\eqref{eq:ho-lagrangian} contains quartic terms in $E^\mu$ and terms
with derivatives of $E^\mu$, makes it rather difficult to obtain
explicit expressions.  

Finally, in the following sections we will discuss the coupling of
tensor supermultiplets to supergravity. In that context it is rather
straightforward to also couple Lagrangians with higher derivatives to
supergravity. As we do not intend to cover this topic in more detail
here, we only present the supergravity coupling to the Lagrangian
\eqref{eq:ho-lagrangian}, restricting ourselves again to the purely
bosonic terms. Such a coupling requires $\mathcal{H}(L)$ to be an
$\mathrm{SU}(2)$ invariant function that is homogeneous of degree
$-2$. The result can then be written as follows,
\begin{equation}
  \label{eq:h-o-lagr-local}
    \mathcal{L} =  \mathcal{L}_1 + \mathcal{L}_2 + \mathcal{L}_3 \,,  
\end{equation}
where 
\begin{equation}
  \label{eq:ho-L-1}
\begin{split}
  e^{-1} \mathcal{L}_1 = & \, \mathcal{H}(L) \Big\{ 
  - \tfrac12 L_{ij} L^{ij} \big( \tfrac{1}{3} R + D \big)^2 
  +\big( E^2 - L^{ij} \mathcal{D}^2 L_{ij} \big) \, \big( \tfrac{1}{3}
  R + D \big)  
  + \vert G \vert^2 \big( \tfrac{1}{6} R + 2 D \big) \\
  & \qquad\quad
  - \mathcal{D}_a E_b \big( {R}^{ab}{}^{i}{}_j(\mathcal{V}) L_{ik}
  \varepsilon^{jk}  
  - \tfrac12 [ T^{ab}{}^{ij} \varepsilon_{ij} \,G  + \mbox{h.c.}]\big)
   \\[1ex] 
  &\qquad\quad  
  +\tfrac1{8} 
  \big( {R}_{ab}{}^{i}{}_j(\mathcal{V}) L_{ik} \varepsilon^{jk} 
  - \tfrac12 [T_{ab}{}^{ij} \varepsilon_{ij} \,G  + \mbox{h.c.}] \big)^2 
  -\tfrac1{64} [T_{ab}{}^{ij} \varepsilon_{ij} G + \mbox{h.c.}]^2 
  \\[1ex]
  &\qquad\quad
  + \vert \mathcal{D}_\mu G \vert^2 
  - \tfrac12 \left( \mathcal{D}^2 L_{ij} \right) \left( \mathcal{D}^2
    L^{ij} \right)    +2 \mathcal{D}_{[a} E_{b]} \, \mathcal{D}^{[a}
  E^{b]}   \Big\} \:,
\end{split}
\end{equation}
\begin{equation}
  \label{eq:ho-L-2}
\begin{split}
   e^{-1} \mathcal{L}_2 =&\, - \tfrac12 \mathcal{H}^{ij}(L) \Big\{ 
  L^{kl} \big( \mathcal{D}_\mu L_{ik} \mathcal{D}^\mu L_{jl} \big)
  \, \big( \tfrac{1}{3} R + D \big) \\[1ex]
  & \qquad\qquad
  + \big(E_b \mathcal{D}_a L_{ij}  +\tfrac12 \mathcal{D}_a L_{ik}
  \mathcal{D}_b L_{jl} \varepsilon^{kl} \big)  
  \big( {R}^{ab}{}^{m}{}_n(\mathcal{V}) L_{mo} \varepsilon^{no} 
  - \tfrac12 [T^{ab}{}^{mn} \varepsilon_{mn} G  + \mbox{h.c.}]
  \big)\\[1ex] 
  &\qquad\qquad
  - \mathcal{D}_\mu L_{ij} \mathcal{D}^\mu \vert G \vert^2 
  - 2 (\mathcal{D}_a L_{ik} \mathcal{D}_b L_{jl} \varepsilon^{kl} -
  E_a \mathcal{D}_b L_{ij}) (\mathcal{D}^a E^b)  
  \\[1ex]
  & \qquad \qquad 
  + \mathcal{D}^2 L_{ij} (\vert G \vert^2 + 2 E^2 )
  + \mathcal{D}^2 L^{kl} ( \mathcal{D}_\mu L_{ik} \mathcal{D}^\mu
  L_{jl} + 2 \varepsilon_{ik} E^\mu \mathcal{D}_\mu L_{jl}  ) 
  \Big\} \;,
\end{split}
\end{equation}
\begin{equation}
  \label{eq:ho-L-3}
\begin{split}
  e^{-1} \mathcal{L}_3 =& \,\tfrac1{4}\mathcal{H}^{ij,kl}(L) \Big\{
  \varepsilon_{ik} \varepsilon_{jl} (\mathcal{D}_\mu L_{mn}
  \mathcal{D}^\mu L^{mn} \vert G \vert^2 
  -  (\vert G \vert^2 + E^2 )^2
  + \tfrac{1}{4} (\mathcal{D}_\mu L_{mn} \mathcal{D}^\mu L^{mn})^2)
  \\[1ex] 
  & \qquad\qquad
  - 2 (\mathcal{D}_\mu L_{ik} \mathcal{D}^{\mu} L_{jl} ) E^2
  + 4 \varepsilon_{ik}[ E^\mu \mathcal{D}_\mu L_{jl}  (\vert G
  \vert^2 + E^2 ) 
  -  ( \mathcal{D}_\mu L_{jm} \mathcal{D}_\nu L_{ln} \varepsilon^{mn}
  ) E^\mu E^\nu] \\[1ex] 
  & \qquad\qquad 
  - \tfrac{1}{2} \varepsilon_{ik} \varepsilon_{jl}  (\mathcal{D}^\mu
  L_{mn} \mathcal{D}^\nu L^{mn})^2  
  - 2 \varepsilon_{ik} E^\nu \mathcal{D}_\mu L_{jl} (\mathcal{D}_\mu
  L_{mn} \mathcal{D}^\mu L^{mn}) 
 \Big\}  \, .  
 \end{split}
\end{equation}
Here $R$ denotes the Ricci scalar associated with the gravitational
field. The other supergravity fields will be introduced in the next
section. Because this expression is based on a chiral superspace
density, one can also introduce elementary vector multiplets as well
as the Weyl multiplet couplings. The latter are accompanied by
additional terms of higher order in the Riemann tensor. Terms such as
these may be important for determining the subleading corrections to
the black hole entropy \cite{LopesCardoso:1998wt}.
%
\section{Coupling to conformal supergravity}
\label{sec:cf-sg-coupling}
\setcounter{equation}{0}
The tensor supermultiplet constitutes also a representation of the
full $N=2$ superconformal algebra \cite{dWvHVP}. In addition to the
translations, Lorentz transformations, and R-symmetry transformations,
the fields are subject to dilatations. In principle, fields also
transform under conformal boosts, but matter multiplets are usually
inert under those. On the fermionic side, the conventional
$Q$-supersymmetry is extended with a second, special, supersymmetry,
called $S$-supersymmetry.

Superconformal transformations can be defined in flat space, with
space-time independent transformation parameters and transformation
rules that explicitly depend on the space-time coordinates. In a
superconformal background, where the translations are replaced by
space-time diffeomorphisms, the transformation rules contain the
various (gauge and other) fields of the superconformal theory. In
their presence the $Q$- and $S$-supersymmetry transformations of the
tensor supermultiplet fields take the following form,
\begin{equation}
  \label{eq:tensor-tr}
  \begin{split}
  \delta L_{ij} =& \,2\,\bar\epsilon_{(i}\varphi_{j)} +2
  \,\varepsilon_{ik}\varepsilon_{jl}\,
  \bar\epsilon^{(k}\varphi^{l)}  \,,\\  
  \delta\varphi^{i} =& \,\Slash{D} L^{ij} \,\epsilon_j +
  \varepsilon^{ij}\,\Slash{\hat E}^I \,\epsilon_j - G \,\epsilon^i
  + 2 L^{ij}\, \eta_j \,,\\  
  \delta G =& \,-2 \,  \bar\epsilon_i \Slash{D} \, \varphi^{i} \,
  - \bar\epsilon_i  ( 6 \, L^{ij} \, \chi_j + \tfrac1{4} \,
    \gamma^{ab}  T_{ab jk} \, \varphi^l \,
    \varepsilon^{ij} \varepsilon^{kl}) + 2 \, \bar{\eta}_i\varphi^{i}
    \, ,\\  
  \delta E_{\mu\nu} =& \, \mathrm{i}\bar\epsilon^i\gamma_{\mu\nu} 
  \varphi^{j} \,\varepsilon_{ij} - \mathrm{i}\bar\epsilon_i\gamma_{\mu\nu}
  \varphi_{j} \,\varepsilon^{ij} \,  + \,2 \mathrm{i} \, L_{ij} \,
  \varepsilon^{jk} \, \bar{\epsilon}^i \gamma_{[\mu} \psi_{\nu ]k}  
  - 2 \mathrm{i}\,  L^{ij} \, \varepsilon_{jk} \, \bar{\epsilon}_i
    \gamma_{[\mu} \psi_{\nu ]}{}^k \, . 
\end{split}
\end{equation}
Here $\epsilon^i$ and $\eta^i$ denote the $Q$- and $S$-supersymmetry
parameters, respectively. The derivatives $D_\mu$ are superconformally
covariant and $\hat E^\mu$ denotes the superconformally covariant
field strength of the tensor field $E_{\mu\nu}$. These quantities,
which will be defined shortly, involve the gauge fields of the
superconformal algebra: the dilatational gauge field $b_\mu$, the
${\rm U}(1)$ and ${\rm SU}(2)$ R-symmetry gauge fields $A_\mu$ and
$\mathcal{V}_\mu{}^i{}_j$, the spin connection field
$\omega_\mu{}^{ab}$, the gauge field $f_\mu{}^a$ associated with
special conformal boosts, and the $Q$- and $S$-supersymmetry gauge
fields $\psi_\mu{}^i$ and $\phi_\mu{}^i$. Not all of these gauge
fields are independent and we refer to the appendix for further
details. Obviously, we also have the vierbein field $e_\mu{}^a$ and
its inverse which are used to convert world to tangent space indices
and vice versa. Apart from the gauge fields, the superconformal theory
contains a complex, selfdual tensor field $T_{ab}{}^{ij}$, a spinor
field $\chi^i$ and a real scalar field $D$, of which only the first
two appear in (\ref{eq:tensor-tr}).

To exhibit some of the details we record the expressions for the
superconformal derivatives and the superconformal tensor field
strength,
\begin{equation}
  \label{eq:cov-qu}
\begin{split}
  D_\mu L_{ij}  = & \, \mathcal{D}_\mu L_{ij} 
  - \bar{\psi}_{\mu(i} \, \varphi_{j)}
  - \varepsilon_{ik} \varepsilon_{jl}\, \bar{\psi}^{(k}_\mu \,
  \varphi^{l)} \,, \\ 
  D_{\mu}\varphi^{i} = & \, \mathcal{D}_\mu\varphi^{i} - 
  L^{ij} \, \phi_{\mu j} 
  - \tfrac{1}{2} \left( \Slash{D}L^{ij} + \varepsilon^{ij}
  \,\Slash{\hat E} \right) \psi_{\mu j} + \tfrac{1}{2} {G}\, 
  \psi_{\mu}{}^i  \, , \\  
  D_{\mu} {G} = & \,\mathcal{D}_\mu{G} - \bar{\phi}_{\mu i}  \varphi^i  
  + \bar{\psi}_{\mu\,i} \, \Slash{D} \varphi^i + 3 \,
  L^{ij}\,\bar{\psi}_{\mu\,i}  \chi_j  
  + \tfrac1{8} \, \bar{\psi}_{\mu i} \, \gamma^{cd} \, T_{cd\,kl} \,
  \varphi_j \, \varepsilon^{ik} \, \varepsilon^{lj} \,,\\
  \hat E^\mu =&\, \tfrac{1}{2}\mathrm{i}\, e^{-1} \, \varepsilon^{\mu \nu
  \rho \sigma}  \Big[\partial_\nu E_{\rho \sigma}   
  - \tfrac{1}{2} \mathrm{i} \bar{\psi}{}^i_\nu  \gamma_{\rho\sigma}
  \varphi^j \varepsilon_{ij} 
  + \tfrac{1}{2}\mathrm{i} \bar{\psi}_{\nu i} \gamma_{\rho\sigma}
  \varphi_{j} \varepsilon^{ij}  
  -\mathrm{i} \, L_{ij}   \varepsilon^{jk} 
  \bar{\psi}_\nu{}^i \gamma_\rho \psi_{\sigma k}\Big] \,.  
\end{split}
\end{equation}
Here the derivatives $\mathcal{D}_\mu$ are covariant with respect to
Lorentz transformations, dilatations and R-symmetry transformations,
\begin{eqnarray}
   \label{eq:cov-s}
   \mathcal{D}_\mu L_{ij} &=& (\partial_\mu - 2 b_\mu ) L_{ij} -
  \cV_{\mu}{}^k{}_{(i} \, L_{j)k} \, \nonumber\\
  \mathcal{D}_\mu\varphi^i &=& ( \partial_\mu - \tfrac1{4}
  \omega_{\mu}{}^{ab} \, \gamma_{ab}  - \tfrac{1}{2}\mathrm{i} A_\mu -
 \tfrac{5}{2} b_\mu)\varphi^{i} + \tfrac{1}{2}
  \cV_{\mu}{}^{i}{}_{j} \, \varphi^j  \,,\nonumber\\
  \mathcal{D}_\mu G &=& ( \partial_\mu - \mathrm{i} \, A_\mu - 3
  \,b_\mu ) G \,.   
\end{eqnarray}

The coupling between a tensor and a reduced chiral supermultiplet is
still possible in a superconformal background \cite{deWit:1982na}. The
reduced chiral multiplet constitutes also a superconformal multiplet
and transforms under $Q$- and $S$-supersymmetry according to
\begin{equation}
  \label{eq:sc-reduced-chiral}
\begin{split}
  \delta X  =& \, \bar\epsilon^i \Omega_i\,,  \\
  \delta \Omega_i =& \, 2\,\Slash{D} X\,\epsilon_i + \tfrac1{2}
  \varepsilon_{ij}\gamma^{\mu\nu} \hat F_{\mu\nu}\,\epsilon^j +
  Y_{ij}\,\epsilon ^j  + 2 X \, \eta_i \,, \\ 
  \delta Y_{ij} = & \, 2\,\bar\epsilon_{(i}\, \Slash{D} \Omega_{j)} +
  2\, \varepsilon_{ik}\varepsilon_{jl}\,\bar\epsilon^{(k} \,\Slash{D}
  \Omega^{l)}  \,. 
\end{split}
\end{equation}
Here we have introduced a superconformal field strength $\hat
F_{\mu\nu}$, defined by 
\begin{eqnarray}
    \label{eq:sc-field-strength}
    \hat F_{\mu\nu} &=& 
    F_{\mu\nu} -  \bar \psi_{[\mu i}\gamma_{\nu]} \Omega_j
    \,\varepsilon^{ij} -\bar \psi_{[\mu}{}^i\gamma_{\nu]} \Omega^j
    \,\varepsilon_{ij}\nonumber\\
    &&{} 
    - X\, \bar \psi_{\mu i} \psi_{\nu j}\,\varepsilon^{ij}  
    - \bar X\, \bar \psi_\mu{}^i \psi_\nu{}^j \,\varepsilon_{ij} 
   - \tfrac{1}{4} \bar{X}\, T_{\mu\nu}{}^{ij} \varepsilon_{ij}-
  \tfrac{1}{4} X \,T_{\mu\nu i j} \varepsilon^{ij} \,,   
\end{eqnarray}
where $F_{\mu\nu}= 2\,\partial_{[\mu}W_{\nu]}$. The supersymmetry
variation of $W_\mu$ remains as given in \eqref{eq:vector-susy}. The
superconformal field strength should be identified with a component
of the superconformal reduced chiral multiplet, as can be seen from its
variation under $Q$- and $S$-supersymmetry,
\begin{equation}
    \label{eq:delta-F-hat}
      \delta\hat F^-_{ab} = \tfrac1{2} \bar \epsilon_i
  \Slash{D}\gamma_{ab} \Omega_j\, \varepsilon^{ij} 
  - \tfrac1{2} \bar \epsilon^i \gamma_{ab}\Slash{D} \Omega^j
    \,\varepsilon_{ij} -\bar \eta_i \gamma_{ab}\Omega_j
  \,\varepsilon^{ij} \,.
\end{equation}
  
The superconformally invariant coupling between the two multiplets is
an extension of (\ref{eq:vector-tensor}),
\begin{eqnarray}
  \label{eq:local-vector-tensor}
  e^{-1}  {\cal L} &=& X\,G + \bar X\,\bar G - \tfrac1{2}
  Y^{ij}\,L_{ij} \nonumber\\
  &&
  - \tfrac1{2} (\bar \psi_\mu{}^i \gamma^\mu \Omega^j
  +\bar X\,\bar \psi_\mu{}^i\gamma^{\mu\nu} \psi_\nu{}^j ) \,
  L_{ij} - \tfrac1{2}( \bar \psi_{\mu i} \gamma^\mu \Omega_j 
  + X\,\bar \psi_{\mu i}\gamma^{\mu\nu} \psi_{\nu j} ) \,   L^{ij}
    \nonumber \\ 
  &&
  + \bar \varphi^i(\Omega_i + X\,\gamma^\mu \psi_{\mu i})  
  + \bar \varphi_i(\Omega^i + \bar X\,\gamma^\mu \psi_\mu{}^i)
    \nn \\
  &&
    - \tfrac1{4}\mathrm{i} \,e^{-1} \varepsilon^{\mu\nu\rho\sigma}
  \,E_{\mu\nu}\, F_{\rho\sigma} \,.    
\end{eqnarray}

Just as in the previous section, we can construct reduced chiral
multiplets from tensor multiplets. Again we start with the complex
scalar $X_I$ defined in (\ref{eq:X}), which transforms into a chiral
spinor $\Omega_{i\,I}$,
\begin{eqnarray}
  \label{eq:X-Omega} 
  X_I &=& {\cal F}_{I,J} \, \bar G^J +
  {\cal F}_{I,JK}{}^{ij}\,\bar\varphi_i{}^J\varphi_j{}^K \,,
  \nonumber \\ 
  \Omega_{i\,I} &=& -2 \,{\cal F}_{I,J}\,\Slash{D}  \varphi_i{}^J
  - {\cal F}_{I,J} (6\,L_{ij}{}^J\chi^j +\tfrac1{4}
  T_{ab}{}^{jk}\,\gamma^{ab}\varphi^{l
  J}\,\varepsilon_{ij}\varepsilon_{kl} ) 
  + 2\, {\cal F}_{I,JKij} \, \bar G^J\, \varphi^{jK} \nonumber\\
  &&
  -  2\, {\cal F}_{I,JK}{}^{kl} \,(\,\Slash{D} L_{ik}{}^J
     -\varepsilon_{ik} \Slash{\hat E}^J ) \varphi_l{}^K   
  + 2\, {\cal F}_{I,JKLij}{}^{kl}  \;\varphi^{jL}
    \,(\bar\varphi_k{}^J\varphi_l{}^K)  \,. 
\end{eqnarray} 
Here the function ${\cal F}_{I,J}(L)$ should again satisfy the
constraints (\ref{eq:chiral-constraints}). But in order that
(\ref{eq:X-Omega}) defines the beginning of a superconformal reduced
chiral multiplet, the component $X_I$ should, in addition, be
invariant under $S$-supersymmetry. This is precisely ensured by the
condition (\ref{eq:sc-tensor}), which implies that the function ${\cal
  F}_{I,J}$ is ${\rm SU}(2)$ invariant and homogeneous of degree $-1$,
so that it has scaling weight $-2$. As it turns out, there are no
further restrictions and we simply record the corresponding
expressions for $Y_{ij\,I}$ and $F_{\mu\nu\,I}$ below, 
\begin{eqnarray}
  \label{eq:sc-Y}
   Y_{ij \,I} &=& -2\,{\cal F}_{I,J}\, \Big[\Box^{\rm c} L_{ij}{}^J +
   3\,D L_{ij}{}^J \Big] 
   -2\,{\cal F}_{I,JKij} \,( \bar G^J\,G^K + \hat E_{\mu}{}^J\,\hat
   E^{\mu K}) \,, 
  \nn\\ 
&&{}
  -2\, {\cal F}_{I,JK}{}^{kl}\,( D_\mu L_{ik}{}^J\,D^\mu
  L_{jl}{}^K + 2\,\varepsilon_{k(i}\, D_\mu L_{j)l}{}^J\,
  \hat E^{\mu K} ) \nn\\ 
&&{} 
   -2 \,{\cal F}_{I,JKLij}{}^{kl} \;\bar\varphi_{k}{}^K\varphi_{l}{}^J 
   \, G^L 
   -2 \,{\cal F}_{I,JKLijkl} \;\bar\varphi^{kK}\varphi^{lJ} \,
   \bar G^L  \nn\\
&&{} 
   + 4\,( {\cal F}_{I,JK m(i}\,\bar\varphi^{mJ} \,\Slash{D}
     \varphi_{j)}{}^K + {\cal F}_{I,JK}{}^{m(k} \bar{\varphi}_m{}^J
   \Slash{D} \varphi^{l)K} \, \varepsilon_{ik} \, \varepsilon_{jl} )
   \nn\\ 
&&{} 
   + 4 \,{\cal F}_{I,JKLn(i}{}^{kl} \,D_\mu L_{j)k}{}^J
     \left(\bar{\varphi}^{nL} \gamma^\mu \varphi_{l}{}^K \right)  \nn\\ 
&&{}
   -4 \,{\cal F}_{I,JKLn(i}{}^{kl}\, \varepsilon_{j)k} \;\left(
   \bar{\varphi}^{nL} \, \Slash{\hat E}^J \, \varphi_{l}{}^K \right) \nn\\
&&{}
   -2 \, {\cal F}_{I,JKLMijmn}{}^{kl} \, \bar\varphi_{k}{}^J
   \varphi_{l}{}^K \;\bar\varphi^{mL} \varphi^{nM}  \nonumber \\
&&{} 
   +12\, {\cal F}_{I,JK k(i}\, L_{j)l}{}^J (\bar\varphi^{kK}\chi^l +
   \varepsilon^{km} \varepsilon^{ln} \bar\varphi_m{}^K\chi_n) \nonumber\\
&&{}
   + \tfrac1{2} {\cal F}_{I,JK k(i} (\varepsilon_{j)m} \,\bar
     \varphi^{kJ}\gamma^{ab} T_{ab}{}^{mn} \varphi^{lK}
     \,\varepsilon_{nl} + \varepsilon^{mk} \,\bar
     \varphi_m{}^{J}\gamma^{ab} T_{ab\,j)n} \varphi_l{}^K
     \,\varepsilon^{nl} )  \;,   \\[2mm] 
\label{eq:sc-eqF}
   F_{\mu\nu\,I} &=& - 2\,{\cal F}_{I,JK}{}^{mn} \,\partial_{[\mu}
   L_{mk}{}^J \, \partial_{\nu]} L_{nl}{}^K \,\varepsilon^{kl} \nn\\ 
&&{} 
-4\, \partial_{[\mu} \left( {\cal F}_{I,J} \,\hat E_{\nu]}{}^J 
+ {\cal F}_{I,JKki}\, \bar \varphi^{kJ}\gamma_{\nu]} \varphi_{j}{}^K
  \,\varepsilon^{ij}\right)  \nn \\ 
&&{}
  + 2\,\partial_{[\mu} \Big( {\cal F}_{I,J} \, \cV_{\nu]}{}^i{}_j \,
    L_{ik}{}^J \, \varepsilon^{jk}  
+ {\cal F}_{I,J} \, \bar{\psi}_{\nu]}{}^i \, \varphi^{jJ} \,
  \varepsilon_{ij} + {\cal F}_{I,J} \, \bar{\psi}_{\nu]i} \,
  \varphi_{j}{}^J \, \varepsilon^{ij} \Big) \, ,
\end{eqnarray} 
where $\Box^{\rm c} L_{ij} = D^aD_a L_{ij}$. An explicit evaluation
leads to the following expression, 
\begin{equation}
  \label{eq:box-L}
\begin{split}
  D_a \, D^a \, L_{ij}{}^I = & \; 
  \mathcal{D}_a D^a L_{ij}{}^{I}  + 2 \,f_{\mu}{}^{\mu} \, L_{ij}{}^{I}
  \\ 
  & \; 
  - \left( \bar{\psi}^{\mu}{}_{\!(i} \, D_\mu \, \varphi_{j)}{}^{I} +
    \varepsilon_{ik} \, \varepsilon_{jl} \, \bar{\psi}^{\mu(k} \, D_\mu
    \varphi^{l)I} \right)  \\
  & \; 
  - \tfrac{1}{16} \, \left( \bar{\varphi}_{(j}{}^{I} \, T^{cd}{}_{i)k}
    \, \gamma_{cd} \, \gamma^\mu \, \psi^k_\mu + \varepsilon_{ik} \,
    \varepsilon_{jl} \, \bar{\varphi}^{(lI} \, T^{k)m cd} \, \gamma_{cd}
    \, \gamma^\mu \, \psi_{\mu m} \right) \\
& \; - \tfrac{3}{2} \, \left( \bar{\psi}_{\mu(i} \, L_{j)k}{}^{I} \,
\gamma^\mu \, \chi^k - \bar{\psi}_\mu{}^k \, \gamma^\mu \, L_{k(i}{}^{I}
\, \chi_{j)} + \bar{\psi}_\mu{}^k \, \gamma^\mu \, \chi_k \, L_{ij}{}^{I}
\, \right) \\
& \; + \tfrac1{2} \left( \bar{\phi}_{\mu(i} \, \gamma^\mu \,
\varphi_{j)}{}^{I} + \varepsilon_{ik} \, \varepsilon_{jl} \,
\bar{\phi}_\mu{}^{\,(k} \, \gamma^\mu \, \varphi^{l)I} \right) \, .
\end{split}
\end{equation}
Obviously the equations \eqref{eq:X-Omega}, \eqref{eq:sc-Y} and
\eqref{eq:sc-eqF} are extensions of the expressions \eqref{eq:X},
\eqref{eq:3.4} and (\ref{eq:Y-F}).  For a single tensor supermultiplet
the results can be compared to \cite{deWit:1982na}. We also note that
it is possible to recast the superconformal extension $\hat F_{abI}$
of \eqref{eq:sc-eqF} in a form where its supercovariance is more manifest,
\begin{equation}
  \label{eq:covF}
\begin{split}
  \hat F_{ab\, I} = & \; -4 \, D_{[a} \, \left( \cF_{I,J} \,\hat E_{b]}^J
 \right)  
 - 2 \, \cF_{I,JK}{}_{ij} \, D_{[a} \, L^{ikJ} \, D_{b]} L^{jlK} \,
 \varepsilon_{kl}  + \cF_{I,J} \, R_{ab}{}^i{}_j(\cV) \, L_{ik}^J \,
 \varepsilon^{jk} \\ 
& \; - \cF_{I,J} \, \bar{\varphi}^{iJ} \, R_{ab}{}^j(Q) \,
 \varepsilon_{ij}  
 - \tfrac{1}{4} \, T_{ab}{}^{ij} \, \varepsilon_{ij} \, \left(
 \cF_{I,J} \, G^{J} + \cF_{I,JK}{}_{kl} \, \bar{\varphi}^{kJ} \,
 \varphi^{lK} \right) \\ 
& \; - \cF_{I,J} \, \bar{\varphi}_i{}^J \, R_{abj}(Q) \,
 \varepsilon^{ij}  
 - \tfrac{1}{4} \, T_{ab ij} \, \varepsilon^{ij} \, \left(
 \cF_{I,J} \, \bar G^{J} + \cF_{I,JK}{}^{kl} \, \bar{\varphi}_k{}^{J} \,
 \varphi_l{}^{K} \right) \\ 
& \;  
  - D_{[a} \left( 4 \, \cF_{I,JK}{}_{ij} \, \bar{\varphi}^{iK} \,
 \gamma_{b]} \, \varphi_k{}^{J} \, \varepsilon^{jk} \right)  \, . 
\end{split}
\end{equation}
However, to derive the Lagrangian it is much more convenient to
work with the expression \eqref{eq:sc-eqF}. 

We now proceed and substitute the above expressions into the supercovariant
density formula \eqref{eq:local-vector-tensor}. In principle this is
straightforward. In doing this we make use of the condition
\eqref{eq:sc-tensor} and in order to express the Lagrangian in terms
of a single function, we also use \eqref{eq:aux-terms}. Dropping a
total derivative term, we then establish that the Lagrangian depends only
on the function $F_{IJ}$ that we encountered earlier in
\eqref{eq:F}. The complete result can then  be presented as follows, 
\begin{equation}
  \label{eq:sc-local-tensor}
 \mathcal{L}_{\mathrm{total}} = e\cL_1 + e\cL_2 + e\cL_3 + e\cL_4 \, , 
\end{equation}
where 
\begin{equation}
  \label{eq:total-L}
  \begin{split}
  \mathcal{L}_1= & \, F_{IJ} \, L_{ij}{}^I \, L^{ijJ} \,
    \Big\{\tfrac{1}{3}\Big[ R + 
     ( e^{-1} \varepsilon^{\mu\nu\rho\sigma}
     \bar{\psi}_\mu{}^i \gamma_\nu \cD_\rho \psi_{\sigma i} 
     - \tfrac{1}{4} \bar{\psi}_\mu{}^i \psi_\nu{}^j\, T^{\mu \nu}{}_{ij} 
          +\mbox{h.c.}) \Big] \\   
     &\qquad 
     +  D + \tfrac{1}{2} (\bar{\psi}_\mu{}^i \gamma^\mu \chi_i
    +\mbox{h.c.}) \Big\}\,, \\
     \mathcal{L}_2 = & \, F_{IJ} \, 
     \Big\{ - \, \tfrac{1}{2} \, \mathcal{D}_\mu L_{ij}{}^{I} \,
     \mathcal{D}^\mu L^{ijJ} + E_{\mu}{}^I \, E^{\mu J}  -
     \bar{\varphi}^{iI} \,\Slash{\mathcal{D}}\varphi_{i}{}^{J} -
     \bar{\varphi}_{i}{}^{I}\,\Slash{\mathcal{D}} \varphi^{iJ} + 
     G^I \bar{G}^J  \\  
   & \qquad 
   - \Big[(\tfrac{1}{8} \bar{\varphi}^{iI} \gamma_{\mu\nu}
   \varphi^{lJ}\,\varepsilon_{ij} \varepsilon_{kl} 
   -\tfrac1{3}  L_{ij}{}^I \,\bar \varphi^{iJ} \gamma_\mu \psi_{\nu k})
   \, T^{\mu\nu jk} +\mbox{h.c.}  \Big] \\ 
   &\qquad 
   - \Big[  L_{ij}{}^I (\tfrac4{3}\bar{\varphi}^{iJ} \gamma^{\mu\nu}
   \mathcal{D}_\mu\psi_\nu{}^j + 2 \bar{\varphi}^{iJ}  \chi^j) 
    +\mbox{h.c.}  \Big] \\
   & 
   \qquad  + \tfrac12 \Big[\bar{\psi}_{\mu i} \,
   [(\Slash{D}+ \Slash{\mathcal{D}}) L^{ijI} - \varepsilon^{ij}
   (\Slash{\hat E}^I+ \Slash{E}^I) ] \,\gamma^\mu\varphi_{j}{}^J  
   +\mbox{h.c.}  \Big] \\
   & \qquad \;
   +     e^{-1} \varepsilon^{\mu\nu\rho\sigma} \bar{\psi}_\nu{}^i \gamma_\rho
   \, \psi_{\sigma k} \,L_{ij}{}^I\mathcal{D}_{\mu}L^{jkJ} - E^{\mu I}
   \, \cV_{\mu}{}^i{}_j \, L_{ik}{}^J \, \varepsilon^{jk} \, ,   \\
  & \qquad \;
  + e^{-1} \varepsilon^{\mu\nu\rho\sigma}\bar{\psi}_\nu{}^i \gamma_\rho
   \, \psi_{\sigma k} \,L_{ij}{}^I\,\varepsilon^{jk} \Big[
   E_\mu{}^J - \tfrac1{4} \bar \psi_\lambda{}^m \gamma_\mu\gamma^\lambda
   \varphi{}^{nJ} \varepsilon_{mn} 
   -\tfrac1{4} \bar \psi_{\lambda m} \gamma_\mu\gamma^\lambda \varphi_n{}^J
   \varepsilon^{mn} \Big]\\
   & \qquad \;
   +\tfrac{1}{4}\,\Big(e^{-1}
   \varepsilon^{\mu\nu\rho\sigma}\bar{\psi}_\nu{}^i \gamma_\rho 
   \, \psi_{\sigma k} \,L_{ij}{}^I\,\varepsilon^{jk}\Big)
   \Big(e^{-1}
   \varepsilon_\mu{}^{\lambda\tau\zeta}\bar{\psi}_\lambda{}^m \gamma_\tau 
   \, \psi_{\zeta} \,L_{mn}{}^J\,\varepsilon^{np}\Big) 
   \Big\} \, ,\\
   \mathcal{L}_3 = & \, \tfrac12 \mathrm{i} e^{-1}
   \varepsilon^{\mu\nu\rho\sigma} \, F_{IJK}{}^{ij}\,E_{\mu\nu}{}^I \,
   \, \partial_\rho L_{ik}{}^J\, \partial_\sigma
   L_{jl}{}^K\,\varepsilon^{kl} \\ 
   & \quad
   + \Big\{ F_{IJKij} \Big(
     \bar{G}^I \, \bar{\varphi}^{iJ} \, \varphi^{jK} + 
     \bar{\varphi}^{iI} \, ( \, \Slash{D} L^{jkJ} \, + \,\varepsilon^{jk}
     \, \Slash{\hat E}^J ) \, \varphi_{k}{}^K \\
   &\qquad\quad  -\bar{\varphi}^{iI}  \varphi^{jJ} \;
   \bar{\psi}_\mu{}^k  \gamma^\mu \varphi_{k}{}^K  
     - \bar{\psi}_\mu{}^i  \varphi^{jI} \; \bar{\varphi}^{kJ}  \gamma^\mu
     \varphi_{k}{}^K   \Big)  + \mbox{h.c.}  \Big\} \,, 
   \\
   \mathcal{L}_{4} = &  \, F_{IJKLij}{}^{kl} 
   \, \bar{\varphi}^{iI}  \varphi^{jJ}  \;  \bar{\varphi}_{k}{}^K
   \varphi_{l}{}^L \,.
  \end{split}
\end{equation}
Setting the fields of the Weyl multiplet to zero, one recovers the
tensor multiplet Lagrangian \eqref{eq:rigid-lagrangian}. For a single
tensor supermultiplet the above expression may be compared to the
result derived in \cite{deWit:1982na}.

\section{Poincar\'e supergravity with tensor multiplets}
\label{sec:Poincare}
\setcounter{equation}{0}
Superconformal matter multiplets coupled to conformal supergravity are
gauge equivalent to matter-coupled Poincar\'e supergravity provided
that enough potential compensating multiplets are present. One
compensating vector multiplet is needed to provide the graviphoton of
$N=2$ Poincar\'e supergravity. For the minimal off-shell versions one
may choose a so-called non-linear multiplet, a hypermultiplet or a
tensor multiplet. In the Poincar\'e context, the conformal symmetries
({\it i.e.}, scale transformations, special conformal boosts and
$S$-supersymmetry) are no longer present. The R-symmetries are usually
absent as well. An exception is the case where a single tensor
multiplet acts as a compensator, because the triplet field $L^{ij}$
has a $\mathrm{U}(1)$ stability subgroup which reflects itself as a
local invariance group of the corresponding Poincar\'e supergravity
Lagrangian \cite{deWit:1982na}. The presence of other multiplets can
nevertheless affect this local invariance, as we shall see in due
course.

In the first subsection we focus on some characteristic features of
the Poincar\'e supergravity Lagrangians with tensor multiplets. As
already explained in section \ref{sec:intro}, it is important to
stress our treatment is based on off-shell multiplets, as it is always
possible to dualize tensor multiplets into hypermultiplets and, in the
presence of suitable isometries, vice versa. This conversion affects,
however, the off-shell structure of the theory. In a second subsection
\ref{sec:tensor-target} we explain the structure of the tensor
multiplet target space. In a third subsection \ref{sec:gf-example} we
work out the example of two tensor multiplets which, upon dualization,
leads to the classification of 4-dimensional quaternion-K\"ahler
manifolds with two abelian isometries. These manifolds include the
so-called universal hypermultiplet which emerges in Calabi-Yau
compactifications of string theory.

\subsection{The general case}
\label{sec:Poincare-general}
In this subsection we discuss the coupling of tensor, vector and
hypermultiplets to supergravity. We first present the Lagrangians in
their superconformally invariant form and exhibit a number of
characteristic features that are relevant in the context of the
super-Poincar\'e formulation. The coupling to tensor multiplets is
based on this paper. For the hypermultiplets we follow the treatment
of \cite{dWKV:1999} and for the vector multiplets we base ourselves on
\cite{DWVP,deWit:1984px} and related references. In all three cases
$n$ will denote the number of independent multiplets. Of course, these
numbers do not have to be equal, but we refrain from introducing extra
notation to make a distinction. As it turns out the couplings for each
of the three types of multiplets can be defined in terms of certain
homogeneous potentials, which we denote by
$\chi_{\mathrm{tensor}}(L)$, $\chi_{\mathrm{hyper}}(\phi)$ and
$\chi_{\mathrm{vector}}(X,\bar X)$, respectively. Under the scale
transformations of the superconformal group, these potentials scale
with weight 2 and they are invariant under the R-symmetry group. As a
result of the scale invariance, the target spaces parametrized by the
scalar fields of each of the three supermultiplets are cones.  For
hypermultiplets the target space is a hyperk\"ahler cone, which is a
cone over a $(4n-1)$-dimensional 3-Sasakian space. The latter is an
$\mathrm{Sp}(1)$ fibration over a $(4n-4)$-dimensional
quaternion-K\"ahler space. The target space of the vector multiplets
is a cone over the product of an $(2n-1)$-dimensional special K\"ahler
space times $S^1$. The target space of the tensor multiplet is the
cone over a $(3n-1)$-dimensional space whose geometrical properties
have not been extensively studied so far.

In the case of tensor and vector multiplets, supersymmetry relates the
gauge fields ({\it i.e.}, the tensor and vector fields) to a special
basis for the scalar fields given by $L_{ij}{}^I$ and $X^\Lambda$,
respectively. The potentials can therefore be generally defined in
terms of these fields. Eventually $L_{ij}{}^I$ and $X^\Lambda$ may be
parametrized in terms of other fields, in which case they will play
the role of sections. The case of hypermultiplets is different in this
respect, because these multiplets do not contain any gauge fields and
have thus no preferred basis for the scalars. Moreover hypermultiplets
do not constitute off-shell supermultiplets, unlike the tensor and
vector supermultiplets. For superconformal hypermultiplets there
exists the so-called hyperk\"ahler potential
$\chi_{\mathrm{hyper}}(\phi)$ \cite{dWKV:1999}, where the fields
$\phi^A$ denote the $4n$ scalar fields corresponding to $n$
hypermultiplets, but there is no a priori definition of the
hyperk\"ahler potential. The fact that we are dealing with hyperk\"ahler
cones implies that the derivative of $\chi_{\mathrm{hyper}}(\phi)$ is
directly related to a homothetic vector denoted by $k^A$,
\begin{equation}
  \label{eq:hk-homothety}
  \frac{\partial\chi_{\mathrm{hyper}}(\phi)}{\partial \phi^A} =
  g_{AB}(\phi)\,k^B(\phi)\,.
\end{equation}
Here $k= k^A \,\partial/\partial\phi^A$ generates the
scale transformations on the target space of the hypermultiplet
scalars and $g_{AB}(\phi)$ denotes the metric on the hyperk\"ahler
cone. We are dealing with an exact homothety, implying that 
\begin{equation}
  \label{eq:exact-homo}
  D_A k^B= \delta_A{}^B\quad \Leftrightarrow \quad
  D_A D_B \chi_{\mathrm{hyper}} = g_{AB}\,,\quad
  \chi_{\mathrm{hyper}}(\phi) = \tfrac12 g_{AB}\, k^Ak^B\,. 
\end{equation}
The covariant derivative contains the Levi-Civit\'a connection
associated with the metric $g_{AB}$.  The formulation of the action
and transformation rules for hypermultiplets is not determined
exclusively in terms of the hyperk\"ahler potential, and we note the
existence of local sections $A_i{}^\alpha(\phi)$ of an
$\mathrm{Sp}(n)\times \mathrm{Sp}(1)$ bundle \cite{Swann} which appear
naturally in the full Lagrangian and transformation rules (here $n$
denotes the number of hypermultiplets and $\alpha=1,\ldots,2n$). Here
group $\mathrm{Sp}(1)$ coincides with the $\mathrm{SU}(2)$ factor of
the R-symmetry group and the $\mathrm{Sp}(n)$ group acts on the
negative-chirality spinors $\zeta^\alpha$ through the indices
$\alpha$. Indices referring to the conjugate $\mathrm{Sp}(n)$
representation will be denoted by $\bar\alpha$ and they label the
positive-chirality spinors $\zeta^{\bar\alpha}$. Under
$S$-supersymmetry the fermions transform into the sections
$A_i{}^\alpha$ mentioned above.

For the three types of supermultiplets, the potentials are defined by
\begin{eqnarray}
  \label{eq:T+V-potential}
  \chi_{\mathrm{tensor}}(L) &=& 2\, F_{IJ} \, L_{ij}{}^I \, L^{ijJ} \,,
  \nonumber\\ 
  \chi_{\mathrm{hyper}}(L) &=& \tfrac1{2} \varepsilon^{ij}
  \bar\Omega_{\alpha\beta} \, A_i{}^\alpha A_j{}^\beta \,,
  \nonumber\\ 
  \chi_{\mathrm{vector}}(X,\bar X) &=&  \mathrm{i} \left( X^\Lambda \bar
  F_\Lambda - \bar X^\Lambda  F_\Lambda\right) = N_{\Lambda\Sigma}
  \,X^\Lambda \bar X^\Sigma\,.
\end{eqnarray}
Here $F_\Lambda$ is the derivative of a holomorphic homogeneous
function $F(X)$ of second degree of the fields $X^\Lambda$ and
$N_{\Lambda\Sigma}$ is defined by 
\begin{equation}
  \label{eq:def-N}
  N_{\Lambda\Sigma} = \frac{\partial^2 \chi_{\mathrm{vector}}(X,\bar
  X) }{\partial X^\Lambda\partial  \bar X^\Sigma}=
  2\,\mathrm{Im}[F_{\Lambda\Sigma}] \,,
\end{equation}
where $F_{\Lambda\Sigma}$ denotes the second derivative of $F(X)$.
Furthermore the symplectic tensor $\bar\Omega_{\alpha\beta}$, which
exists for any hyperk\"ahler space, can be defined as follows,
\begin{equation}
  \label{eq:Omega-hk}
  \bar\Omega_{\alpha\beta}= \tfrac1{2} \varepsilon_{ij}\,  g_{AB} 
  \,\gamma^{A i}{}_{\alpha} \,\gamma^{B j}{}_{\beta} \,,  
\end{equation}
where $\gamma^{A i}{}_\alpha$ denotes a generalized vielbein that
converts hyperk\"ahler target-space indices into $\mathrm{Sp}(n)\times
\mathrm{Sp}(1)$ indices. This quantity appears in the supersymmetry
transformations of the hypermultiplet scalars,
\begin{equation}
  \label{eq:susy-phi}
  \delta\phi^A = 2\,
  (\gamma^A{}_{i\bar\alpha}\,\bar\epsilon^i\zeta^{\bar\alpha} +
  \gamma^{Ai}{}_\alpha \, \bar\epsilon_i\zeta^{\alpha})\,. 
\end{equation}

The presentation above shows that the combined target space is a
product of three cones, each with its own potential. The potentials
satisfy properties that are very similar to \eqref{eq:hk-homothety}
and \eqref{eq:exact-homo}, except that there is no need to use
covariant derivatives as in \eqref{eq:exact-homo}.  For tensor
multiplets the corresponding equations are given by
\eqref{eq:homothetic}, \eqref{eq:chi-metric} and
\eqref{eq:tensor-potential-2}. The homogeneity of the potentials
follows from 
\begin{eqnarray}
  \label{eq:homo-potential}
  L_{ki}{}^I \,\frac{\partial\chi_{\mathrm{tensor}}(L)}{\partial
  L_{kj}{}^I} &=& \tfrac1{2} \delta^j{}_i \,\chi_{\mathrm{tensor}}(L)
  \,, \nonumber\\[1ex] 
   k^A(\phi) \,\frac{\partial\chi_{\mathrm{hyper}}(\phi)}{\partial
  \phi^A} &=& 2\, \chi_{\mathrm{hyper}}(\phi)\,, \nonumber\\[1ex]
  X^\Lambda\, \frac{\partial\chi_{\mathrm{vector}}(X,\bar X)}{\partial
  X^\Lambda} = \bar X^\Lambda\, \frac{\partial\chi_{\mathrm{vector}}(X,\bar
  X)}{\partial\bar X^\Lambda}  &=& \chi_{\mathrm{vector}}(X,\bar
  X) \,. 
\end{eqnarray}
For the tensor and vector multiplet potentials the above equations
also imply the invariance under R-symmetry. For the hyperk\"ahler
potential the equations for R-symmetry involve the relevant Killing
vectors, or the complex structures of the hyperk\"ahler cone.

Let us now exhibit some characteristic terms of the three Lagrangians,
and compare them (eventually we also consider the sum of the three
Lagrangians), 
\begin{eqnarray}
  \label{eq:list}
  e^{-1}\mathcal{L}_{\mathrm{tensor}} &=&{} 
  \tfrac{1}{6}\,\chi_{\mathrm{tensor}} \,
  \Big[ R + 
     ( e^{-1} \varepsilon^{\mu\nu\rho\sigma}
     \bar{\psi}_\mu{}^i \gamma_\nu \cD_\rho \psi_{\sigma i} 
     - \tfrac{1}{4} \bar{\psi}_\mu{}^i \psi_\nu{}^j\, T^{\mu \nu}{}_{ij} 
          +\mbox{h.c.}) \Big] \nn \\   
  &&{}
     + \tfrac{1}{2}\,\chi_{\mathrm{tensor}} \,\,\Big[
     D + \tfrac{1}{2} (\bar{\psi}_\mu{}^i \gamma^\mu \chi_i
    +\mbox{h.c.}) \Big]  \,, \nn \\[1ex]  
    &&{} - \tfrac{1}{2}  F_{IJ} \, \mathcal{D}_\mu L_{ij}{}^{I}
     \, \mathcal{D}^\mu L^{ijJ} \nn \\[1ex]    
   &&{} - \Big(\,\frac{\partial\chi_{\mathrm{tensor}}}{\partial
    L^{ijI}} \,   
    \Big[\tfrac2{3}\bar{\varphi}^{iI} \gamma^{\mu\nu}
   \mathcal{D}_\mu\psi_\nu{}^j +  \bar{\varphi}^{iI} \chi^j 
   - \tfrac1{6}\,\bar \varphi^{iI} \gamma_\mu \psi_{\nu k} \, 
    T^{\mu\nu jk}  \Big]
    +\mbox{h.c.}\Big) \;, \nn \\[1ex]  
  e^{-1}\mathcal{L}_{\mathrm{hyper}} &=&{} 
  \tfrac{1}{6}\,\chi_{\mathrm{hyper}} \,  \Big[ R + 
     ( e^{-1} \varepsilon^{\mu\nu\rho\sigma}
     \bar{\psi}_\mu{}^i \gamma_\nu \cD_\rho \psi_{\sigma i} 
     - \tfrac{1}{4} \bar{\psi}_\mu{}^i \psi_\nu{}^j\, T^{\mu \nu}{}_{ij} 
          +\mbox{h.c.}) \Big] \nn \\[1ex]  
  &&{}
     + \tfrac{1}{2}\,\chi_{\mathrm{hyper}}\,\Big[
     D + \tfrac{1}{2} (\bar{\psi}_\mu{}^i \gamma^\mu \chi_i
    +\mbox{h.c.}) \Big]  \,, \nn \\[1ex]  
    &&{} - \tfrac{1}{2}  g_{AB}\,\mathcal{D}_\mu \phi^A
    \,\mathcal{D}^\mu\phi^B  \nn \\[1ex]  
   &&{} 
   - \frac{\partial\chi_{\mathrm{hyper}}}{\partial \phi^A}  \Big(
   \gamma^A{}_{i\bar\alpha} 
    \Big[\tfrac2{3}\bar{\zeta}^{\bar\alpha} \gamma^{\mu\nu}
   \mathcal{D}_\mu\psi_\nu{}^i + \bar{\zeta}^{\bar\alpha}   \chi^i 
   - \tfrac1{6}  \,\bar \zeta^{\bar\alpha} \gamma_\mu \psi_{\nu j} 
   \, T^{\mu\nu ij}\Big]
    +\mbox{h.c.}  \Big) \;, \nn \\[1ex]  
  e^{-1}\mathcal{L}_{\mathrm{vector}} &=&{} 
   \tfrac{1}{6} \chi_{\mathrm{vector}} \,
  \Big[ R + 
     ( e^{-1} \varepsilon^{\mu\nu\rho\sigma}
     \bar{\psi}_\mu{}^i \gamma_\nu \cD_\rho \psi_{\sigma i} 
     +\tfrac{1}{2} \bar{\psi}_\mu{}^i \psi_\nu{}^j\, T^{\mu \nu}{}_{ij} 
          +\mbox{h.c.}) \Big] \nn \\[1ex]   
  &&{}
     - \chi_{\mathrm{vector}} \,\Big[
     D + \tfrac{1}{2} (\bar{\psi}_\mu{}^i \gamma^\mu \chi_i
    +\mbox{h.c.}) \Big]  \,, \nn \\[1ex]  
    &&{} - N_{\Lambda\Sigma} \, \mathcal{D}_\mu
     X^{\Lambda}\,\mathcal{D}^\mu \bar X^{\Sigma}
     \nn \\[1ex]  
   &&{} - \Big(\frac{\partial\chi_{\mathrm{vector}}}{\partial X^\Lambda}   
    \Big[\tfrac1{3}\bar\Omega_i{}^{\Lambda} 
    \gamma^{\mu\nu} \mathcal{D}_\mu\psi_\nu{}^i - 
    \bar{\Omega}_i{}^\Lambda \chi^i + \tfrac1{6} \bar
    \Omega_i{}^\Lambda \gamma_\mu \psi_{\nu j}\, T^{\mu\nu ij}   \Big] 
    +\mbox{h.c.} \Big)   \;. 
\end{eqnarray}
 
The equations \eqref{eq:list} exhibit a rather uniform structure for
the various couplings. Especially the couplings of tensor multiplets
and hypermultiplets is closely related, which is not surprising in
view of the fact that the tensor multiplets can be dualized to
hypermultiplets. The fact that the potentials for the tensor multiplet
cones and the hyperk\"ahler cones are identical, a result derived at
the end of subsection \ref{sec:tensor-cones}, makes the agreement even
more close. 

With the vector multiplet there are subtle differences reflected in
the relative coefficients. It is well known that these differences are
crucial for converting to the Poincar\'e formulation. The above
Lagrangians still contain gauge degrees of freedom associated with
certain superconformal symmetries. The symmetry under conformal boosts
is manifest. Because only the dilatational gauge field $b_\mu$
transforms under this symmetry, it follows that the Lagrangians are
independent of $b_\mu$, as can be verified by explicit computation.
The dilatational symmetry is still intact and we can impose a corresponding
gauge condition. The obvious condition is to set the coefficient of
the Ricci scalar in the combined Lagrangian to a constant, {\it i.e.},
\begin{equation}
  \label{eq:D-gauge}
  \tfrac1{6}\chi_{\mathrm{tensor}} + \tfrac1{6}
  \chi_{\mathrm{hyper}}+ \tfrac1{6} \chi_{\mathrm{vector}}=  -
  \frac{1}{2\,\kappa^2} \,, 
\end{equation}
so that we end up with a conventional Einstein-Hilbert term. Observe
that, in order to describe scalar fields with kinetic terms of the
correct sign, it follows that the cone metrics can not be positive
definite. Under $Q$-supersymmetry the condition \eqref{eq:D-gauge} is
not invariant, and it is convenient to exploit $S$-supersymmetry to
set its variation to zero by a second gauge choice. This motivates the
condition,
\begin{equation}
  \label{eq:S-gauge}
  2\,\frac{\partial\chi_{\mathrm{tensor}}}{\partial
    L^{ijI}} \,\varphi^{jI} 
  + 2\, \frac{\partial\chi_{\mathrm{hyper}}}{\partial \phi^A}
    \gamma^A{}_{i\bar\alpha}\, \zeta^{\bar\alpha} 
  + \frac{\partial\chi_{\mathrm{vector}}}{\partial X^\Lambda}
    \,\Omega_i{}^\Lambda = 0 \,.   
\end{equation}
Concentrating on the second and fourth lines of the three Lagrangians
\eqref{eq:list} we see that the fields $D$ and $\chi^i$ act as
Lagrange multipliers, which leads, when combined with the above gauge 
choices, to the following results,
\begin{eqnarray}
  \label{eq:DS-gauge}
  \chi_{\mathrm{tensor}} + \chi_{\mathrm{hyper}} &=& - 2\,\kappa^{-2} 
  \,, \nn\\
  \chi_{\mathrm{vector}}&=&  - \kappa^{-2} \,, \nn\\
  \frac{\partial\chi_{\mathrm{tensor}}}{\partial
    L^{ijI}} \;\varphi^{jI} 
  + \frac{\partial\chi_{\mathrm{hyper}}}{\partial \phi^A}\;
    \gamma^A{}_{i\bar\alpha}\, \zeta^{\bar\alpha}  &=& 0\,,\nn \\
  \frac{\partial\chi_{\mathrm{vector}}}{\partial X^\Lambda}
    \;\Omega_i{}^\Lambda &=& 0 \,.    
\end{eqnarray}
Because we used the field equations corresponding to one bosonic and
eight fermionic fields, supersymmetry is no longer realized off
shell. As a result of the above procedure the sum of the Lagrangians
\eqref{eq:list} reduces to 
\begin{eqnarray}
  \label{eq:gf-L}
  e^{-1}\mathcal{L}_{\mathrm{combined}} &=&{} 
  -\frac1{2\,\kappa^2} R -\frac1{2\,\kappa^2}
  \Big[e^{-1}\varepsilon^{\mu\nu\rho\sigma} 
     \bar{\psi}_\mu{}^i \gamma_\nu \cD_\rho \psi_{\sigma i} 
     - \tfrac{1}{4} \bar{\psi}_\mu{}^i \psi_\nu{}^j\, T^{\mu \nu}{}_{ij} 
          +\mbox{h.c.} \Big] \nn \\[1ex]  
  &&{} - \tfrac{1}{2}  F_{IJ} \, \mathcal{D}_\mu L_{ij}{}^{I}
     \, \mathcal{D}^\mu L^{ijJ} 
    - \tfrac{1}{2}  g_{AB}\,\mathcal{D}_\mu \phi^A
    \,\mathcal{D}^\mu\phi^B  
    - N_{\Lambda\Sigma} \, \mathcal{D}_\mu
     X^{\Lambda}\,\mathcal{D}^\mu \bar X^{\Sigma} \,. 
\end{eqnarray}
In this formulation the scalar fields are constrained by the first two
equations of \eqref{eq:DS-gauge} and corresponding restrictions exist
on the fermions. The full action is now invariant under general
coordinate transformations, local Lorentz transformations, local
supersymmetry (defined as a field-dependent linear combination of $Q$-
and $S$-supersymmetry) and local R-symmetry. Clearly the
hypermultiplet and tensor multiplet fields are entangled whereas the
vector multiplet fields remain separate, a feature that has been
known for some time.

\subsection{The tensor multiplet target space}
\label{sec:tensor-target}
We now specialize to the tensor scalars $L_{ij}{}^I$ and analyze their
corresponding target space. It is convenient to change notation at
this point and rescale the fields $L_{ij}{}^I$ by the inverse of
$\chi_{\mathrm{tensor}}$ so that the $L_{ij}{}^I$ are scale invariant
(we refrain from imposing a gauge condition). The rescaled fields are
then constrained to a hypersurface,
\begin{equation}
  \label{eq:hypersurface-L}
  2\, F_{IJ}(L) \,L_{ij}{}^I L^{ijJ} = 1 \,.
\end{equation}
Furthermore we use a vector notation for the fields $L_{ij}{}^I$,
according to
\begin{equation}
  \label{eq:new-L}
  L_{ij}{}^I = -\mathrm{i}\, \vec L^I\cdot(\vec \sigma)^k{}_i\,
  \varepsilon_{jk}  \,, 
\end{equation}
where $\vec \sigma = (\sigma_1,\sigma_2,\sigma_3)$ are the Pauli
matrices (with $\sigma_1\sigma_2\sigma_3=\mathrm i$) so that
$L_{ij}{}^I L^{ij J} = 2\, \,\vec L^I\cdot\vec 
L^J$. With these definitions we find,
\begin{equation}
  \label{eq:4]kinetic-L}
  \tfrac1{2} F_{IJ}(L)\, \mathcal{D}_\mu L_{ij}{}^I \mathcal{D}^\mu
    L_{ij}{}^I  =  \frac{(\partial_\mu
    \chi_{\mathrm{tensor}})^2}{4\,\chi_{\mathrm{tensor}}} +  
    \chi_{\mathrm{tensor}}\, F_{IJ}(L)\, \mathcal{D}_\mu \vec L^I \cdot 
    \mathcal{D}^\mu \vec L^J  \,, 
\end{equation}
which shows that we are indeed dealing with a cone over a
$(3n-1)$-dimensional space parametrized by the constrained coordinates
$\vec L^I$.

Let us now write the bosonic terms of the Lagrangian
\eqref{eq:total-L} in terms of the rescaled variables,
\begin{eqnarray}
  \label{eq:tensor-boson}
  e^{-1} \mathcal{L}_{\mathrm{tensor}}  &=& 
  \chi_{\mathrm{tensor}} \Big[ \tfrac{1}{6} R  
     +  \tfrac1{2} D  -\tfrac1{4} (\partial_\mu
    \ln \chi_{\mathrm{tensor}})^2 \Big] \nn\\[1ex]
    &&{} 
    - \chi_{\mathrm{tensor}} 
    F_{IJ}(L)\; (\partial_\mu\vec L^I- \vec{\mathcal{V}}_\mu\times \vec
    L^I) \cdot  (\partial^\mu\vec L^J -\vec{\mathcal{V}}^\mu\times \vec 
    L^J)   \nonumber\\[1ex]
    &&{}
    + \chi_{\mathrm{tensor}}^{-1} F_{IJ}(L) \Big[E_{\mu}{}^I \, E^{\mu J} + 
     G^I \bar{G}^J\Big] \nn\\[1ex]
    &&{}
    + 2\,F_{IJ}(L)\, E^{\mu I} \, \vec L^J\, \vec \cV_{\mu}  
    -\tfrac12 \mathrm{i} e^{-1}
   \varepsilon^{\mu\nu\rho\sigma} \, \vec F_{IJK}(L) \cdot
  (\partial_\rho \vec L^I \times \partial_\sigma \vec L^J)
  \,E_{\mu\nu}{}^K \,, 
\end{eqnarray}
where $\vec F_{IJK} =\partial F_{IJ}/\partial  \vec L^K$  and 
\begin{equation}
  \label{eq:vec-V}
  \mathcal{V}_\mu{}^i{}_j = \mathrm{i} \vec{\mathcal{V}}_\mu \, (\vec
  \sigma)^i{}_j\,.   
\end{equation}
To eliminate the auxiliary $\mathrm{SU}(2)$ gauge fields
$\vec {\mathcal{V}}_\mu$, the matrix that multiplies the terms 
quadratic in these fields is relevant,
\begin{equation}
  \label{eq:Mmat}
  \left[ \mathbf{M} \right]^{rs} =  F_{IJ} \, \vec{L}^I \cdot
  \vec{L}^J \, \delta^{rs} - L^{Ir} \, F_{IJ} \, L^{Js} \,,   
\end{equation}
where $r,s =1,2,3$ denote vector indices. It is clear that this matrix
has zero eigenvalues whenever all vectors $\vec L^I$ are aligned,
which is related to the fact that these configurations leave a
subgroup of $\mathrm{SU}(2)$ invariant. This is especially relevant
for the case of a single tensor multiplet, which always leaves a
subgroup invariant, so that the approach sketched below is not
applicable. For several tensor multiplets generic configurations
correspond to matrices $\mathbf{M}$ with non-vanishing determinant. In
that case the equations of motion for $\vec{\mathcal{V}}_\mu$ lead to
\begin{equation}
  \label{eq:sol-V}
  \vec{\mathcal{V}}_\mu = {\mathbf{M}}^{-1} \,( 
  \vec{L}^I \times \partial_\mu\vec{L}^J + \chi_{\mathrm{tensor}}^{-1}\,
  E_{\mu}{}^I \, \vec{L}^J) \,F_{IJ}\,,
\end{equation}
where the inverse $\mathbf{M}^{-1}$ equals\footnote{
  Direct verification of this result makes use of the identity for
  general $3\times3$ matrices $\mathcal{O}$,
  \begin{equation}
    \label{eq:O3}
  \mathcal{O}^3 - \mathrm{tr}(\mathcal{O})\,\mathcal{O}^2 + \tfrac1{2}
  [(\mathrm{tr}(\mathcal{O}))^2 - \mathrm{tr}(\mathcal{O}^2)]\, 
  \mathcal{O}  = \det(\mathcal{O})\;   \mathbf{1}\,. \nonumber 
\end{equation}
   } 
\begin{equation}
  \label{eq:inverse}
  \left[\mathbf{M}^{-1} \right]_{rs} = \frac{1}{\det(\mathbf{M})}
  \left[ \tfrac12 
  F_{IJ} F_{KL} (\vec{L}^I \times \vec{L}^K ) \cdot ( \vec{L}^J
  \times \vec{L}^L ) \, \delta_{rs} 
  + L_r{}^{I} \,  F_{IK}( \vec{L}^K \cdot \vec{L}^L ) F_{LJ}\,L_s{}^{J}
  \right] \, .  
\end{equation}
The determinant of $\mathbf{M}$ is given by 
\begin{equation}
  \label{eq:det-M}
  \det(\mathbf{M}) = \tfrac1{3} (F_{IJ}\vec L^I \cdot \vec L^J)^3 -
  \tfrac1{3} F_{IJ}(\vec L^J \cdot \vec L^K) F_{KL}(\vec L^L \cdot \vec
  L^M) F_{MN}(\vec L^N \cdot \vec L^I)  \,. 
\end{equation}

The Lagrangian \eqref{eq:tensor-boson} is invariant under tensor gauge
transformations, up to a surface term. The latter originates
exclusively from the last term in \eqref{eq:tensor-boson}. To
establish this one needs to use the condition
\begin{equation}
  \label{eq:laplace-vec}
  \frac{\partial}{\partial \vec L^I} \cdot \frac{\partial}{\partial
  \vec L^J}\; F_{KL}=0\,,
\end{equation}
which follows from \eqref{eq:chi-metric}, and which was extensively
discussed in section \ref{sec:rigid-tensors}. Under local
$\mathrm{SU}(2)$ transformations the Lagrangian is also invariant up
to a surface term. These transformations can be written as
\begin{equation}
  \label{eq:su2-vec}
  \delta \vec L^I = \vec\Lambda\times \vec L^I\,,\qquad
  \delta\vec{\mathcal{V}}= \partial_\mu\vec\Lambda + \vec\Lambda\times
  \vec{\mathcal{V}} \,, 
\end{equation}
where $\vec\Lambda(x)$ represents the infinitesimal space-time
dependent parameters of $\mathrm{SU}(2)$, and the variation of the
Lagrangian (resulting from the last two terms in
\eqref{eq:tensor-boson}) reads,
\begin{equation}
  \label{eq:var-su2-L}
  \delta_\Lambda \mathcal{L}_{\mathrm{tensor}} =
  \partial_\mu\Big(-\mathrm{i} 
  \varepsilon^{\mu\nu\rho\sigma} \,  F_{IJ} \, \vec
  L^I\cdot\partial_\nu\vec\Lambda \, E_{\rho\sigma}{}^J\Big)\,. 
\end{equation}

Substituting \eqref{eq:sol-V} in the Lagrangian
\eqref{eq:tensor-boson} then leads to the following Lagrangian 
\begin{eqnarray}
  \label{eq:tensor-boson-noV}
  e^{-1} \mathcal{L}_{\mathrm{tensor}}  &=& 
  \chi_{\mathrm{tensor}} \Big[ \tfrac{1}{6} R  
     +  \tfrac1{2} D  -\tfrac1{4} (\partial_\mu
    \ln \chi_{\mathrm{tensor}})^2 \Big] \nn\\[1ex]
    &&{} 
    - \chi_{\mathrm{tensor}}\;
    \Big[\mathcal{G}^{(1)}_{IJ} \,\partial_\mu\vec L^I \,
    \partial^\mu\vec L^{J} + 
    \mathcal{G}^{(2)}_{IJ,KL} \, (\vec L^I\cdot\partial_\mu \vec L^J) \, 
    (\vec L^K\cdot\partial_\mu \vec L^L) \Big] 
    \nonumber\\[1ex]
    &&{}
    + \chi_{\mathrm{tensor}}^{-1}\,\Big[ \mathcal{H}^{(1)}_{IJ}\,
    E_{\mu}{}^I \, E^{\mu J} 
     +  F_{IJ} \,G^I \bar{G}^J \Big]  \nn\\[1ex]
    &&{}
    + E^{\mu I} \, \mathcal{H}^{(2)}_{IJ}\, \vec L^J\cdot(\vec
    L^K\times\partial_\mu\vec L^L) \, F_{KL} \nonumber\\[1ex] 
    &&{}
    -\tfrac12 \mathrm{i} e^{-1} 
   \varepsilon^{\mu\nu\rho\sigma} \, \vec F_{IJK} \cdot
  (\partial_\rho \vec L^I \times \partial_\sigma \vec L^J)
  \,E_{\mu\nu}{}^K \,, 
\end{eqnarray}
where 
\begin{eqnarray}
  \label{eq:FHG}
    \mathcal{H}^{(1)}_{IJ} &= & F_{IJ} + F_{IK} \, L_r{}^{K} \,
    (\Mm^{-1})^{rs} \, L_s{}^{L} \, F_{LJ}\,, \nn \\ 
    \mathcal{H}^{(2)}_{IJ} &=& \frac1{\det(\mathbf{M})}
    \Big[ [(F_{KL}\vec L^K\cdot\vec L^L)^2 - F_{KL}\vec L^L\cdot\vec
    L^M F_{MN}\vec L^N\cdot\vec L^K] \,F_{IJ} \nn \\
    &&{} \qquad \qquad
    + 2\,F_{IK}\vec L^K\cdot\vec L^L F_{LM}\vec L^M\cdot\vec L^N
    F_{NJ} \Big] \,,  \nn  \\  
    \mathcal{G}^{(1)}_{IJ} &= & F_{IJ} - \frac{(F_{KL}\vec
    L^K\cdot\vec L^L)^2 + F_{KL}\vec L^L\cdot\vec L^M F_{MN}\vec
    L^N\cdot\vec L^K}{2\,\det(\mathbf{M})}
    \; F_{IP}\vec L^P\cdot\vec L^Q F_{QJ}  \nonumber\\
    &&{}
    + \frac1{\det(\mathbf{M})}  F_{IK}\vec L^K \cdot\vec L^L
    F_{LM}\vec L^M \cdot\vec L^N   F_{NP}\vec L^P \cdot\vec L^Q
    F_{QJ}    \,, \nonumber\\
    \mathcal{G}^{(2)}_{IJ,KL} &= & \frac1{\det(\mathbf{M})} \; 
    F_{IM}\vec L^M \cdot\vec L^N F_{NK}\, 
    F_{JP}\vec L^P\cdot\vec L^Q F_{QL}\nn\\
    &&{}
    + \frac{(F_{KL}\vec L^K\cdot\vec L^L)^2 + F_{KL}\vec L^L\cdot\vec
    L^M F_{MN}\vec L^N\cdot\vec L^K}{2\,\det(\mathbf{M})} 
    \; F_{IL}\;  F_{JK}    \nn\\
    &&{}
    -\frac1{\det(\mathbf{M})} \;\Big[  F_{IM}\vec L^M \cdot\vec L^N
    F_{NP}\vec L^P \cdot\vec L^Q F_{QL}\; F_{JK}+ (I\leftrightarrow K;
    J\leftrightarrow L) \Big]\,.
\end{eqnarray}
The elimination of the $\mathrm{SU}(2)$ gauge fields does not affect
the invariance under local $\mathrm{SU}(2)$. This means that the
target space involves only $3(n-1)$ scalar fields, subject to the
constraint \eqref{eq:hypersurface-L}, which in the present notation
reads, 
\begin{equation}
  \label{eq:hypersurface-L-2}
  F_{IJ}\, \vec L^I\cdot\vec L^J = \tfrac1{4}\,.
\end{equation}

In principle one can now construct the most general variety of these
spaces, starting from the (homogeneous) potential
$\chi_{\mathrm{tensor}}$ written in terms of $\mathrm{SU}(2)$
invariant variables. Subsequently one imposes the conditions
\eqref{eq:homothetic} and \eqref{eq:chi-metric}, which yield a number
of second-order differential equations. Every solution of these
equations yields a corresponding Lagrangian. Finally one imposes the
constraint \eqref{eq:hypersurface-L-2}. At this point one has the
option to convert the tensor fields $E_{\mu\nu}{}^I$ to scalars and
obtain a quaternion-K\"ahler manifold of dimensions $4(n-1)$ with $n$
abelian isometries. We already mentioned that the case $n=1$ is
special, and also the cases $n=2$ and 3 are rather specific. The $n>3$
cases can be dealt with in a more generic way. In the next subsection
we demonstrate this procedure for the case of $n=2$ tensor multiplets.
In this way we will rather conveniently obtain the classification of
4-dimensional quaternion-K\"ahler spaces with two commuting isometries
presented in \cite{CP:2001}. We intend to return to an analysis of the
higher-$n$ cases in the future.

\subsection{The case of two tensor supermultiplets}
\label{sec:gf-example}
To illustrate the procedure sketched in the previous subsection we
start by considering the most general potential $\chi_{\rm tensor}$
for two tensor multiplets, $\vec L^1$ and $\vec L^2$. This potential
must be invariant under $\mathrm{SU}(2)$ rotations and homogeneous of
first degree under a uniform rescaling of the $L_{ij}{}^I$. In order
to incorporate these constraints it is convenient to introduce the
SU(2) invariant variables,
\begin{equation}
  \label{eq:reduced-suv}
  s = \vec L^1\cdot\vec  L^1 \; , 
  \qquad 
  u =  \frac{(\vec L^1 \cdot\vec L^1)\,(\vec L^2 \cdot \vec L^2)  -
  (\vec L^1 \cdot\vec L^2)^2 }{s^2}  \; , 
  \qquad 
  v =  \frac{ \vec L^1 \cdot\vec L^2 }{s} \,.
\end{equation}
Note that $s,u\geq0$ and that $u$ vanishes whenever the two vectors
$\vec L^1$ and $\vec L^2$ are aligned. For $u=0$ we thus expect
singularities as this value corresponds to field configurations that
are invariant under a subgroup of $\mathrm{SU}(2)$.  When expressed in
terms of the above variables, the most general potential must be of the
form
\begin{equation}
  \label{eq:chiA}
  \chi_{\rm tensor} = \sqrt{2s} \, f(u,v) \, . 
\end{equation}
Substituting this ansatz into \eqref{eq:homothetic} determines
the entries of the matrix $F_{IJ}$ to be
\begin{equation}
  \label{eq:FIJ2t}
  F_{IJ} = \frac{1}{\sqrt{2 s}} \, 
  \begin{pmatrix}
   \tfrac12 f - v f_v - u f_u + v^2 f_u & \tfrac1{2} f_v -vf_u \\[2mm]
   \tfrac1{2} f_v -vf_u &  f_u   
\end{pmatrix}\,. 
\end{equation}
We also need the $2\times 2$ matrix
\begin{equation}
  \label{eq:LL-2t}
  \vec L^I\cdot \vec L^J =  s \, 
  \begin{pmatrix}
   1  & v  \\[2mm]
   v  & u+v^2    
\end{pmatrix}\,. 
\end{equation}
Imposing the constraint \eqref{eq:chi-metric} leads to the following
partial differential equation for $f(u,v)$, 
\begin{equation}
  \label{eq:2Tpdgl1}
  f_{vv} + 4 u\, f_{uu} = 0 \,.  
\end{equation}
Thus the most general Lagrangian for two tensor multiplets coupled to
supergravity is based on the potential \eqref{eq:chiA} with the
function $f(u,v)$ subject to \eqref{eq:2Tpdgl1}.  In passing, we note
the perturbatively corrected hypermultiplet
\cite{Antoniadis:2003sw,Anguelova:2004sj} corresponds to the following
expression for the underlying tensor multiplet potential,
\begin{equation}
  \label{eq:pert-universal}
  \chi_{\rm tensor} = - 2 \, \sqrt{s}(u +  2 c) \,, 
\end{equation}
which indeed satisfies the differential equation
\eqref{eq:2Tpdgl1}. Here the constant $c$ is determined by the one-loop
string correction to the universal hypermultiplet. 
 
For a small number of tensor multiplets the various terms in the
bosonic Lagrangian are most conveniently obtained from
\eqref{eq:tensor-boson}. Since we already established the invariance
under local $\mathrm{SU}(2)$ we can consider a special gauge. In
principle the $\mathrm{SO}(3)$ vector space can be decomposed into the
two-dimensional space spanned by the vectors $\vec L^I$ and a
one-dimensional subspace orthogonal to it. By adopting a gauge
condition one can ensure that the derivatives $\partial_\mu\vec L^I$
take their values in the subspace spanned by the $\vec L^I$. With this
condition one derives that $\vec L^I\cdot(\vec L^J\times\vec L^K)=0$,
and this suffices to show that the last term of
\eqref{eq:tensor-boson} vanishes. Likewise, upon substituting the
expression \eqref{eq:sol-V} for the $\mathrm{SU}(2)$ gauge field, all
terms linear in $E^{\mu\,I}$ vanish as well for the same reason.
Note that the above considerations pertain specifically to the case
$n=2$.

Now let us be more explicit about the gauge choice. By an appropriate
rotation we can bring the $\vec L^I$ in the form 
\begin{equation}
  \label{eq:gauge-L}
  \vec L^1 = (\sqrt{s},0,0)\,,\qquad 
    \vec L^2 = (v \sqrt{s}, \sqrt{su},0 )\,,
\end{equation}
so that their inner products satisfy \eqref{eq:LL-2t}. Because the
$\mathrm{SU}(2)$ is local we can ensure that this decomposition holds
for all space-time points, so that the $\partial_\mu\vec L^I$ can be
obtained consistently from \eqref{eq:gauge-L}. It is now easy to
evaluate the matrix $\mathbf{M}$ defined in \eqref{eq:Mmat}, which has
a block-diagonal decomposition,
\begin{equation}
  \label{eq:M-gauge}
  \mathbf{M} = \sqrt{\frac s{8}} 
  \begin{pmatrix} Q_{2\times 2} & 0 \\[3mm]
    0& f \end{pmatrix} \;,
\end{equation}
with the $2\times 2$ matrix $Q$ defined by 
\begin{equation}
  \label{eq:def-Q}
  Q  = \begin{pmatrix} 2uf_u& -\sqrt{u} f_v \\[3mm]
     -\sqrt{u} f_v & f- 2u f_u \end{pmatrix} \;.
\end{equation}
The fields $\vec{\mathcal{V}}_\mu$ can now be evaluated explicitly and
substituted into the Lagrangian.  This leads to the following kinetic
terms for the scalar fields $s,u$ and $v$,
\begin{equation}
  \label{eq:scalar-kin}
  \begin{split}
  \mathcal{L}_{\mathrm{scalars}} =
   - \frac{e\, \chi_{\mathrm{tensor}}}{\sqrt{2s}} \Big[&
   \frac{f}{8s} \,(\partial_\mu s)^2 + \frac{1}{2} \partial_\mu
  s\,\partial^\mu f  +\frac{s\,f_u}{4u} \Big((\partial_\mu u)^2
  +4u(\partial_\mu v)^2  \Big) \\
  &\, - \frac s{8 u f} (f_v\,
  \partial_\mu u - u f_u \, \partial_\mu v )^2 \Big] \,.
  \end{split}
\end{equation}
Taking into account the fact that the three fields $s,u,v$ are
constrained by \eqref{eq:hypersurface-L-2}, which implies
\begin{equation}
  \label{eq:constraint-s}
  s = \frac1{2\,f^2} \,,
\end{equation}
one directly establishes 
\begin{equation}
  \label{eq:n=2-Lag}
\begin{split}
  e^{-1} \mathcal{L}_{n=2}  =& \,  
  \chi_{\mathrm{tensor}} \Big[ \tfrac{1}{6} R  
     +  \tfrac1{2} D  -\tfrac1{4} (\partial_\mu
    \ln \chi_{\mathrm{tensor}})^2 \Big] \\[1ex] 
    &\,{} 
   - \frac{\chi_{\mathrm{tensor}}\, \det (Q) }{(4\, u f)^2}\; \Big[
   (\partial_\mu u)^2 +   4u\, (\partial_\mu v)^2\Big] \\[1ex]
   &\,{}
    + \chi_{\mathrm{tensor}}^{-1}\,\Big[ \mathcal{H}^{(1)}_{IJ}\,
    E_{\mu}{}^I \, E^{\mu J} 
     +  F_{IJ} \,G^I \bar{G}^J \Big] \,, 
\end{split}
\end{equation}
where 
\begin{equation}
  \label{eq:H-function}
  \mathcal{H}^{(1)}_{IJ} = \frac{ f\, \det(Q)} {u\, \sqrt{8\,s}} \,
  [N\, Q^{-2} N^{\rm T}]_{IJ} \,. 
\end{equation}
Here $s$ is determined by \eqref{eq:constraint-s} and the matrix $N$
is defined by 
\begin{equation}
  \label{eq:matrix-N}
  N=\begin{pmatrix} \sqrt{u} & -v\\[2mm] 0 & 1
  \end{pmatrix} \;. 
\end{equation}

It is straightforward to perform a duality transformation by
introducing Lagrange multipliers $\phi_I$ to impose the Bianchi
identity on the field strengths $E^{\mu I}$ and by subsequently
integrating out the field strengths. The resulting line element is
then equal to (we suppress the overall factor
$\chi_{\mathrm{tensor}}$),
\begin{equation}
  \label{eq:line-element}
  \mathrm{d}s^2 = \frac{\det (Q) }{(4\, u f)^2}\; [
   \mathrm{d}u^2 +   4u\, \mathrm{d}v^2] +
   [\mathcal{H}^{(1)}]^{IJ}\,\mathrm{d}\phi_I \,\mathrm{d}\phi_J \,,
\end{equation}
where $[\mathcal{H}^{(1)}]^{IJ}$ is the inverse of
\eqref{eq:H-function}. Upon a change of coordinates this line element
coincides precisely with the expression derived by Calderbank and
Pedersen for the general class of selfdual Einstein metrics with two
commuting Killing fields \cite{CP:2001}. In this work the matrices $Q$
and $N$ are related but not quite identical to the matrices used
above. Hence, the formalism discussed in this paper enables a
straightforward and elegant derivation of this classification.

\section*{Acknowledgement}
We thank Stefan Vandoren for valuable discussions and for his
participation during an early stage of this work. F.S. is supported by
a European Commission Marie Curie Postdoctoral Fellowship under
contract number MEIF-CT-2005-023966.  This work is partly supported by
NWO grant 047017015, EU contracts MRTN-CT-2004-005104 and
MRTN-CT-2004-512194, and INTAS contract 03-51-6346. 

\begin{appendix}
%
\section{Superconformal calculus}
\label{App:SC}
\setcounter{equation}{0}
Throughout this paper we use Pauli-K\"all\'en conventions and follow
the notation used {\it e.g.} in \cite{BH}. Space-time indices are
denoted by $\mu,\nu,\ldots$ and Lorentz indices by $a,b,\ldots$.
Furthermore $\mathrm{SU}(2)$-indices are denoted by $i,j,\ldots$ and
the corresponding $\mathrm{SO}(3)$-indices by $r,s,\ldots$. All
\mbox{(anti-)}sym\-metrizations are with unit strength. Majorana
spinors are defined by $\bar{\varphi} = \varphi^T \, C$, where the
four-dimensional charge conjugation matrix $C$ satisfies
\begin{equation}
  \label{eq:1c-conj}
  - \gamma_\mu^{\rm T} = C \, \gamma_\mu \, C^{-1} \, , \qquad
  \gamma_5^{\rm T} = C \, \gamma_5 \, C^{-1} \, , \qquad C^{\rm T} = - C
  \,.
\end{equation}

The superconformal algebra consists of general coordinate, local
Lorentz, dilatation, special conformal, chiral $\mathrm{U}(1)$ and
$\mathrm{SU}(2)$, and $Q$- and $S$-supersymmetry transformations. 
Under Q-supersymmetry, S-supersymmetry and conformal
transformations the independent fields of the Weyl multiplet transform
as follows: 
\begin{equation}
  \label{eq:sc-transf}
\begin{split}
\delta e_\mu{}^a & = \bar{\epsilon}^i \, \gamma^a \psi_{i \mu} +
\bar{\epsilon}_i \, \gamma^a \psi_{ \mu}^i \, , \\ 
\delta \psi_{\mu}{}^{i} & = 2 \,\cD_\mu \epsilon^i - \tfrac{1}{8}
 T_{ab}{}^{ij} \gamma^{ab}\gamma_\mu \epsilon_j - \gamma_\mu \eta^i \,
, \\ 
\delta b_\mu & = \tfrac{1}{2} \bar{\epsilon}^i \phi_{\mu i} -
\tfrac{3}{4} \bar{\epsilon}^i \gamma_\mu \chi_i - \tfrac{1}{2}
\bar{\eta}^i \psi_{\mu i} + \mbox{h.c.} + \Lambda^a_K e_{\mu a} \, ,
\\ 
\delta A_{\mu} & = \tfrac{1}{2} i \bar{\epsilon}^i \phi_{\mu i} +
\tfrac{3}{4} i \bar{\epsilon}^i \gamma_\mu \, \chi_i + \tfrac{1}{2} i
\bar{\eta}^i \psi_{\mu i} + \mbox{h.c.} \, , \\  
\delta \cV_\mu{}^{i}{}_j & = 2\, \bar{\epsilon}_j \phi_\mu{}^i - 3
\bar{\epsilon}_j \gamma_\mu \, \chi^i + 2 \bar{\eta}_j \, \psi_{\mu}{}^i
- (\mbox{h.c. ; traceless}) \, , \\   
\delta T_{ab}{}^{ij} & = 8 \,\bar{\epsilon}^{[i} R_{ab}{}^{j]}(Q) \, ,
\\ 
\delta \chi^i & = - \tfrac{1}{12} \gamma^{ab} \, \Slash{D} T_{ab}{}^{ij}
\, \epsilon_j + \tfrac{1}{6} R(\mathcal{V})_{\mu\nu}{}^i{}_j
  \gamma^{\mu\nu} \epsilon^j -
\tfrac{1}{3} \mathrm{i} R_{\mu\nu}(A) \gamma^{\mu\nu} \epsilon^i + D
\epsilon^i + 
\tfrac{1}{12} \gamma_{ab} T^{ab ij} \eta_j \, , \\ 
\delta D & = \bar{\epsilon}^i \,  \Slash{D} \chi_i +
\bar{\epsilon}_i \,\Slash{D}\chi^i \, . 
\end{split}
\end{equation}
Here $\Lambda^a_K$ is the transformation parameter for conformal
transformations.  The full superconformally covariant derivative is
denoted by $D_\mu$ while $\cD_\mu$ denotes a covariant derivative with
respect to Lorentz, dilatation, chiral $\mathrm{U}(1)$, and
$\mathrm{SU}(2)$ transformations, \emph{e.g.}, (also see eqs.\ 
\eqref{eq:cov-qu} and \eqref{eq:cov-s})
\begin{equation}
  \label{eq:D-epslon}
\cD_{\mu} \epsilon^i = \left( \partial_\mu - \tfrac{1}{4}
\omega_\mu{}^{cd} \, \gamma_{cd} + \tfrac1{2} \, b_\mu -
\tfrac{1}{2}\mathrm{i} \, A_\mu  \right) \epsilon^i + \tfrac1{2} \,
\cV_{\mu}{}^i{}_j \, \epsilon^j  \,.  
\end{equation}

The supercovariant curvature tensors\footnote{
  We corrected a typo  in \cite{BH} in the definition of 
  $R_{\mu\nu}(D)$.}
used here as well as in the main part of the paper are defined as 
\begin{equation}
  \label{eq:curvatures}
\begin{split}
  R_{\mu \nu}{}^a(P)  = & \, 2 \, \partial_{[\mu} \, e_{\nu]}{}^a + 2 \,
  b_{[\mu} \, e_{\nu]}{}^a -2 \, \omega_{[\mu}{}^{ab} \, e_{\nu]b} -
  \tfrac1{2} ( \bar\psi_{[\mu}{}^i \gamma^a \psi_{\nu]i} +
  \mbox{h.c.} ) \, , \\ 
  R_{\mu \nu}{}^i(Q) = & \, 2 \, \cD_{[\mu} \psi_{\nu]}{}^i -
  \gamma_{[\mu}   \phi_{\nu]}{}^i - \tfrac{1}{8} \, T^{abij} \,
  \gamma_{ab} \, \gamma_{[\mu} \psi_{\nu]j} \, , \\ 
  R_{\mu \nu}(A) = & \, 2 \, \partial_{[\mu} A_{\nu ]} - i \left( \tfrac12
    \bar{\psi}_{\mu}{}^i \phi_{\nu]i} + \tfrac{3}{4} \bar{\psi}_{[\mu}{}^i
    \gamma_{\nu ]} \chi_i - \mbox{h.c.} \right) \, , \\
  R_{\mu \nu}{}^i{}_j(\cV){} =& \, 2\, \partial_{[\mu} \cV_{\nu]}{}^i{}_j +
  \cV_{[\mu}{}^i{}_k \, \cV_{\nu]}{}^k{}_j  +  2 (
    \bar{\psi}_{[\mu}{}^i \, \phi_{\nu]j} - \bar{\psi}_{i[\mu} \,
    \phi_{\nu]}{}^j ) 
  -3 ( \bar{\psi}_{[\mu}{}^i \gamma_{\nu]} \chi_j -
    \bar{\psi}_{[\mu j} \gamma_{\nu]} \chi^i ) \\ 
& \, - \delta_j{}^i ( \bar{\psi}_{[\mu}{}^k \, \phi_{\nu]k} -
  \bar{\psi}_{[\mu k} \, \phi_{\nu]}{}^k ) 
  + \tfrac{3}{2}\delta_j{}^i (\bar{\psi}_{[\mu}{}^k \gamma_{\nu]}
  \chi_k - \bar{\psi}_{[\mu k} \gamma_{\nu]} \chi^k)  \, , \\ 
  R_{\mu \nu}{}^{ab}(M) = & 
  \, 2 \,\partial_{[\mu} \omega_{\nu]}{}^{ab} - 2 \omega_{[\mu}{}^{ac}
  \omega_{\nu]c}{}^b  
  - 4 f_{[\mu}{}^{[a} e_{\nu]}{}^{b]} 
  + \tfrac12 (\bar{\psi}_{[\mu}{}^i \, \gamma^{ab} \,
  \phi_{\nu]i} + \mbox{h.c.} ) \\ 
& \, + ( \tfrac12 \bar{\psi}_{[\mu}{}^i \, T^{ab}_{ij} \,
  \phi_{\nu]}{}^j  
  - \tfrac{3}{4} \bar{\psi}_{[\mu}{}^i \, \gamma_{\nu]} \, \gamma^{ab}
  \chi_i 
  - \bar{\psi}_{[\mu}{}^i \, \gamma_{\nu]} \,R^{ab}{}_i(Q)  
  + \mbox{h.c.} ) \, , \\
  R_{\mu \nu}(D) = & \,2\,\partial_{[\mu} b_{\nu]} - 2 f_{[\mu}{}^a 
  e_{\nu]a}  
  - \tfrac{1}{2} \bar{\psi}_{[\mu}{}^i \phi_{\nu]i} + \tfrac{3}{4}
    \bar{\psi}_{[\mu}{}^i \gamma_{\nu]} \chi_i 
    - \tfrac{1}{2} \bar{\psi}_{[\mu i} \phi_{\nu]}{}^i + \tfrac{3}{4}
  \bar{\psi}_{[\mu i} \gamma_{\nu]} \chi^i \,. 
\end{split}
\end{equation}
The remaining curvature tensors, $R_{\mu\nu}{}^i(S)$ and
$R_{\mu\nu}{}^a(K)$, are not needed here, but may be found in
\cite{BH}. There are three conventional constraints,
\begin{equation}
  \label{eq:phi-f}
\begin{split}
  & R_{\mu \nu}(P) =  0 \, , \\
  & \gamma^\mu \left( R_{\mu \nu}(Q)^i + \tfrac12 \gamma_{\mu \nu}
  \chi^i \right) = 0 \, , \\ 
  & 
  e^{\nu}{}_b \,R_{\mu \nu}(M)_a{}^b - \I \tilde{R}_{\mu a}(A) +
  \tfrac1{8} T_{abij} T_\mu{}^{bij} -\tfrac{3}{2} D \,e_{\mu a} = 0
  \,, 
\end{split}
\end{equation}
which determine the fields $\omega_{\mu}{}^{ab}$, $\phi_\mu{}^i$ and 
$f_{\mu}{}^a$. We only used the expressions,  
\begin{equation}
  \label{eq:A.4}
\begin{split}
  \phi_\mu{}^i & \, = \tfrac12 \left( \gamma^{\rho \sigma} \gamma_\mu -
    \tfrac{1}{3} \gamma_\mu \gamma^{\rho \sigma} \right) \left( \cD_\rho
    \psi_\sigma{}^i - \tfrac{1}{16} T^{abij} \gamma_{ab} \gamma_\rho
    \psi_{\sigma j} + \tfrac{1}{4} \gamma_{\rho \sigma} \chi^i \right) \,
  ,  \\ 
  f_\mu{}^{\mu} & \, = \tfrac{1}{6} R - D - \left( \tfrac1{12} e^{-1}
    \varepsilon^{\mu \nu \rho \sigma} \bar{\psi}_\mu{}^i \, \gamma_\nu
    \cD_\rho \psi_{\sigma i} - \tfrac1{12} \bar{\psi}_\mu{}^i
    \psi_\nu{}^j T^{\mu \nu}{}_{ij} - \tfrac1{4} \bar{\psi}_\mu{}^i
    \gamma^\mu \chi_i + 
    \mbox{h.c.} \right) \, .  
\end{split}
\end{equation}
When combining the conventional constraints with the various Bianchi
identities one establishes that the curvatures are not all
independent. For instance we note the relation,
\begin{equation}
\label{eq:R(A-D)} 
     \tilde{R}_{\mu\nu}(D) - \mathrm{i} R_{\mu\nu}(A) = \, 0  \, .
\end{equation} 

For convenience, the Weyl and chiral weights together with the 
chirality of the spinors belonging to the Weyl, tensor and vector
multiplet, are summarized in the tables \ref{one}, \ref{two} and
\ref{three}, respectively.
 

%
\begin{table}[p]
\begin{tabular*}{\textwidth}{@{\extracolsep{\fill}} |c||cccccccc|ccc||ccc| } 
\hline 
 & &\multicolumn{9}{c}{Weyl multiplet} & &
 \multicolumn{2}{c}{parameters} & \\  \hline \hline 
 field & $e_\mu{}^{a}$ & $\psi^i_\mu$ & $b_\mu$ & $A_\mu$ &
 $\cV_\mu{}^i{}_j$ & $T^{ij}_{ab} $ & 
 $ \chi^i $ & $D$ & $\omega_\mu^{ab}$ & $f_\mu{}^a$ & $\phi^i_\mu$ &
 $\epsilon^i$ & $\eta^i$  
 & \\ \hline
$w$  & $-1$ & $-\tfrac12 $ & 0 &  0 & 0 & 1 & $\tfrac{3}{2}$ & 2 & 0 &
1 & $\tfrac12 $ & $ -\tfrac12 $  & $ \tfrac12  $ & \\ \hline 
$c$  & $0$ & $-\tfrac12 $ & 0 &  0 & 0 & $-1$ & $-\tfrac{1}{2}$ & 0 &
0 & 0 & $-\tfrac12 $ & $ -\tfrac12 $  & $ -\tfrac12  $ & \\ \hline 
 $\gamma_5$   &  & + &   &    &   &   & + &  &  &  & $-$ & $ + $  & $
 -  $ & \\ \hline 
\end{tabular*}
\renewcommand{\baselinestretch}{1}
\parbox[c]{\textwidth}{\caption{\label{one}{\footnotesize
Weyl and chiral weights ($w$ and $c$, respectively) and fermion
chirality $(\gamma_5)$ of the Weyl multiplet component fields and the
supersymmetry transformation parameters.}}} 
\end{table}
%
\begin{table}[p]
\begin{center}
\begin{tabular*}{8.5cm}{@{\extracolsep{\fill}}|c||ccccc| } 
\hline 
 & & \multicolumn{3}{c}{Tensor multiplet} & \\  \hline \hline
 field & $E_{\mu \nu} $ & $L^{ij}$ & $\varphi_i$ & $G$ & $\cFm_{IJ}$  \\ \hline
$w$  & $0$ & $ 2 $ & $\tfrac{5}{2}$ &  3 & -2  \\ \hline
$c$  & $0$ & $ 0 $ & $- \tfrac12$ &  1 & 0   \\ \hline
$\gamma_5$   &  &  &  $ - $ &    &    \\ \hline
\end{tabular*}
\renewcommand{\baselinestretch}{1}
\parbox[c]{8.5cm}{\caption{\label{two}{\footnotesize
Weyl and chiral weights ($w$ and $c$, respectively) and fermion
chirality $(\gamma_5)$ of the tensor multiplet component fields.}}} 
\end{center}
\end{table}
\begin{table}[p]
\begin{center}
\begin{tabular*}{8.5cm}{@{\extracolsep{\fill}}|c||cccc| } 
\hline 
 & & \multicolumn{2}{c}{Vector multiplet} & \\  \hline \hline
 field & $X $ & $\Omega_i$ & $W_\mu$ & $Y_{ij}$   \\ \hline
$w$  & $1$ & $ \tfrac{3}{2} $ & 0 &  2   \\ \hline
$c$  & $-1$ & $- \tfrac12$ & 0 &  0   \\ \hline
$\gamma_5$   &  & $ + $ &   &       \\ \hline
\end{tabular*}
\renewcommand{\baselinestretch}{1}
\parbox[c]{8.5cm}{\caption{\label{three}{\footnotesize
Weyl and chiral weights ($w$ and $c$, respectively) and fermion
chirality $(\gamma_5)$ of the vector multiplet component fields.}}} 
\end{center}
\end{table}
\clearpage 
\end{appendix}


\end{document}